\documentclass[preprintnumbers,aps,prd,floatfix,nofootinbib,onecolumn]{revtex4}
\usepackage{qcd}
\usepackage{epsfig}
\usepackage{graphicx}
\usepackage{url,hyperref}
\usepackage{bm}
\usepackage{float}
\usepackage{orcidlink}

\newcommand{\mspec}{M_{\text{F}}}
\newcommand{\mospec}{M_{\text{0F}}}
\newcommand{\mspecff}{M_{\text{D}}}
\newcommand{\mospecff}{M_{\text{0D}}}

\begin{document}

\title{The resolution to the problem of consistent large transverse momentum in TMDs}

\preprint{JLAB-THY-23-3771}

\author{J.~O.~Gonzalez-Hernandez \orcidlink{0000-0003-4352-9564}\,}
\email{joseosvaldo.gonzalezhernandez@unito.it}
\affiliation{Dipartimento di Fisica, Universit\`a degli Studi di Torino, Via P. Giuria 1, I-10125, Torino, Italy}
\affiliation{INFN, Sezione di Torino, Via P. Giuria 1, Torino, I-10125, Italy}
%
\author{T.~Rainaldi \orcidlink{0000-0002-8342-6765}\,}
\email{train005@odu.edu}
\author{T.~C.~Rogers \orcidlink{0000-0002-0762-0275}\,}
\email{trogers@odu.edu}
\affiliation{Department of Physics, Old Dominion University, Norfolk, VA 23529, USA}
\affiliation{Jefferson Lab, 12000 Jefferson Avenue, Newport News, VA 23606, USA}

\begin{abstract}
Parametrizing TMD parton densities and fragmentation functions in ways that consistently match their large transverse momentum behavior in standard collinear factorization has remained notoriously difficult. We show how the problem is solved in a recently introduced set of steps for combining perturbative and nonperturbative transverse momentum in TMD factorization. Called a ``bottom-up'' approach in a previous article, here we call it a ``hadron structure oriented'' (HSO) approach to emphasize its focus on preserving a connection to the TMD parton model interpretation. We show that the associated consistency constraints improve considerably the agreement between parametrizations of TMD functions and their large-$k_T$ behavior, as  calculated in collinear factorization. The procedure discussed herein will be important for guiding future extractions of TMD parton densities and fragmentation functions and for testing TMD factorization and universality. We illustrate the procedure with an application to semi-inclusive deep inelastic scattering (SIDIS) structure functions at an input scale $Q_0$, and we show that there is improved consistency between different methods of calculating  at moderate transverse momentum. We end with a discussion of plans for future phenomenological applications. 
\end{abstract}

\date{\today}

\maketitle

\section{Introduction}
\label{s.intro}

Transverse momentum dependent (TMD) parton distribution functions (pdfs) and/or fragmentation functions (ffs), together with the TMD factorization theorems~\cite{Collins:1981uk,Collins:1984kg,Collins:2011qcdbook}, have acquired a wide range of applications in hadronic, nuclear and high energy phenomenology~\cite{Angeles-Martinez:2015sea,anselmino2016topical} over the past few decades. They are useful both for studying the role of intrinsic or nonperturbative effects in hadrons~\cite{Gao:2018grv,Bressan:2018mtd} and for predicting transverse momentum distributions in cross sections after evolution to high energies. In the former case, they play an important role in testing, and thus refining, the partonic constituent interpretation of hadron structure.
However, separating truly nonperturbative or intrinsic transverse momentum effects from the perturbatively generated transverse momentum that is calculable with collinear factorization has remained a difficult challenge.  It is a problem that limits the predictive power of TMD factorization and creates ambiguity about the interpretation of phenomenologically extracted nonperturbative objects. This is especially the case with lower invariant energies, near the boundary between what may be considered an appropriate hard scale. 

To see the issues clearly, recall that one normally categorizes contributions to a TMD cross section, such as semi-inclusive deep inelastic scattering (SIDIS), according its transverse momentum regions. On one hand, the small transverse momentum regions are associated with nonperturbative effects in hadronic bound states. There, purely nonpertubative parton model descriptions are often quite successful phenomenologically, especially for moderate $Q$. On the other hand,  in the regions of large $\Tsc{q}{}$ perturbative tails where $\Tsc{q}{} \approx Q$,  calculations can be performed in fixed order perturbation theory with collinear factorization with $Q$ as a hard scale. One example of this way of cataloging physically distinct regions can be seen in the treatment of 
SIDIS in Ref.~\cite{COMPASS:2017mvk}, following the theoretical work in Ref.~\cite{Anselmino:2006rv}, where Fig.~17 shows two separate fits for the small transverse momentum nonperturbative peak and the large transverse momentum perturbative tail. Reference~\cite{COMPASS:2017mvk} attributes the behavior in each region to a different underlying physical mechanism, namely a nonperturbative peak and a perturbative tail at small and large transverse momentum respectively. 
There one reads that ``the two exponential functions in our parameterisation $F_1$ can be attributed to two completely different underlying physics mechanisms that overlap in the region
$P_{hT} \simeq 1.0 \, \text{(GeV/c)}^2$.''

Individual TMD pdfs and ffs can be viewed in an analogous way.
When the transverse momentum in an individual TMD pdf is comparable to the renormalization scale $\mu$, $\Tsc{k}{} \approx \mu \approx \sqrt{\zeta}$, it is straightforward to calculate the TMD pdf  directly from its operator definition at a fixed, low order in collinear factorization. This provides a very useful consistency check in phenomenological implementations. Namely, the parametrizations of TMD pdfs and ffs that are used in phenomenology must, within perturbative or power-suppressed errors, match their expressions as obtained from fixed order collinear factorization in the large transverse momentum ($\Tsc{k}{} \approx \mu$) limit as $\mu \to \infty$.  

However, most implementations of TMD phenomenology from the past decade find tension between the extracted TMD functions and their large transverse momentum limits as calculated in fixed order collinear factorization. Consider, for instance, the far right panel in Fig.~6 of \cite{Boglione:2014oea}. The pale blue dot-dashed curve is the cross section calculation performed with TMD pdfs and ffs (the so-called ``$W$ term'' or ``TMD term''). 
This is to be compared with the dashed green curve (the ``asymptotic'' term), which represents the large transverse momentum asymptote of the cross section, calculated theoretically in collinear factorization.  In principle, consistency demands that the TMD term and the asymptotic term approximately overlap in a range of $\Lambda_\text{QCD} \ll \Tsc{q}{} \ll Q$. As the figure illustrates, this is not the case, at least for calculations done with standard parametrizations of collinear and TMD functions. It is only at the extremely high energies, shown in the far left plot, that a region starts to emerge where the asymptotic and TMD terms (very roughly) begin to overlap at intermediate transverse momentum. While the exact details of the mismatch depend on the specifics of the implementation, the trend appears to be quite general \cite{Nadolsky:1999kb,Echevarria:2018qyi,Moffat:2019pci,Bacchetta:2019tcu}, and it applies to other processes where TMD factorization is often used\footnote{A successful implementation of the matching, that predates modern TMD factorization theorems, was presented in  \cite{Arnold:1990yk}}. The overall picture suggests that elements are still missing from the standard way that TMD factorization gets implemented at a practical level. 

A separate issue is that, for transverse momentum comparable to the hard scale ($\Tsc{q}{} \approx Q$), the small $\Tsc{q}{} \ll Q$ approximation fails and a so-called ``$Y$-term'' is needed in order to get an accurate cross section calculation. However, the consistency problems alluded to above appear already at the level of the $\Tsc{q}{} \ll Q$ contribution.  In past papers, this small-$\Tsc{q}{}$ contribution has sometimes been called the ``$W$-term,'' and it is the contribution that involves TMD correlation functions. It, and the TMD correlation functions from which it is composed, is the main focus of this paper. 
Throughout this paper, we will call it the ``TMD term'' to emphasize its connection to TMD pdfs and ffs. 

In this paper, we will show how to recover consistency between the TMD term and the large-$\Tsc{q}{}$ asymptote by using an approach recently introduced by two of us~\cite{Gonzalez-Hernandez:2022ifv}. In the process, we will diagnose some of the complications that, in the past, have been responsible for a mismatch. One problem arises from the way one imposes constraints of the form 
\begin{equation}
\label{e.int_rel_basic}
f_{i/p}(x) \approx \int \diff{^2 \T{k}{}}{} f_{i/p}(x,\T{k}{}) \, ,
\end{equation}
where here there is an ``$\approx$'' rather than a strict equality because such integrals are generally ultraviolet (UV) divergent and are only satisfied literally in a strict parton model interpretation where the pdf is a literal probability density. To maintain a partonic interpretation, one hopes to preserve an approximate version of \eref{int_rel_basic} as accurately as possible. 
For a given parametrization of $f_{i/p}(x)$, the parameters in a model of the nonperturbative transverse momentum in $f_{i/p}(x,\T{k}{})$ are constrained by \eref{int_rel_basic}. Now, in standard procedures for implementing the Collins-Soper-Sterman (CSS) formalism and similar approaches to TMD factorization, the nonperturbative transverse momentum dependence is contained within transverse coordinate space functions that are usually labeled $g_{i/p}(x,\T{b}{})$ (and $g_K(\T{b}{})$ for the Collins-Soper (CS) kernel). To our knowledge, however, constraints corresponding to \eref{int_rel_basic} are never directly imposed upon the $g_{i/p}(x,\T{b}{})$ functions in phenomenological applications that use the $g$-function approach. As explained in Ref.~\cite{Gonzalez-Hernandez:2022ifv}, this will in general produce  mismatches between the models of nonperturbative transverse momentum and the collinear functions $f_{i/p}(x)$ that are used to describe the perturbative tails. 
We will see with explicit examples in this paper that the effects of the mismatch can propagate in transverse momentum space and spoil the matching at intermediate regions of transverse momentum. Although we will mainly use standard $\msbar$ collinear pdfs and ffs for the parts of calculations that require collinear factorization, we will sometimes find it convenient in intermediate steps to work with collinear pdfs and ffs defined as the transverse momentum integrals of TMD pdfs and ffs with UV cutoffs, 
\begin{equation}
f^c(x;\mu) \equiv \pi \int_0^{\mu^2} \diff{\Tscsq{k}{}}{} f_{i/p}(x,\T{k}{};\mu;\zeta) \, , \label{e.cutoff_def}
\end{equation}
where $\mu$ is the usual auxiliary mass parameter associated with $\msbar$ renormalization and $\zeta$ is the CS scale. The ``$c$'' superscript on the left-hand side stands for ``cutoff scheme.'' As will be explained in the text, the cutoff-defined and $\msbar$ pdfs and ffs can be translated into one another at large $\mu$ via relatively simple perturbative correction terms, so the choice of which one to use is ultimately largely a matter of convenience. However, the explicit expressions for \eref{cutoff_def} do 
have the advantage of a natural and direct connection to a TMD parton model interpretation. 

A coherent treatment of the issues discussed above will be necessary in order for a meaningful analysis of future SIDIS data in terms of TMD parton correlation functions to be possible, and for the interpretability of, for example, forthcoming results from the CEBAF $12$~GeV program~ or a $24$~GeV upgrade~\cite{Arrington:2021alx}, as well as for a future electron-ion collider (EIC). In Ref.~\cite{Gonzalez-Hernandez:2022ifv} we called the treatment a ``bottom-up'' approach to distinguish it from more conventional treatments whose starting points were tailored to very high energies. In this paper we will instead call it the ``hadron structure oriented'' (HSO) approach to emphasize the central role of the nonperturbative input and the focus on preserving a partonic interpretation. 

In this paper, we will set up the calculation of the TMD term for SIDIS using the HSO approach of \cite{Gonzalez-Hernandez:2022ifv}, and we will analyze in detail the transition to the large $\Tsc{q}{}$ asymptotic term. 
We will show how imposing the integral relation in \eref{cutoff_def}, ensuring a smooth transition between nonperturbative TMD behavior at small transverse momentum and the large transverse momentum tails, and several other adjustments to the conventional treatment fixes the problems outlined above. Specifically, we will show how to ensure that nonperturbative TMD pdf and ff parameterizations remain reasonably consistent with their expected large transverse momentum behavior, especially near the input scale. This work complements other efforts to address similar problems, for example \cite{Qiu:2000hf,Grewal:2020hoc} imposes continuity and smoothness conditions on $g$-functions directly in coordinate space.

The structure of the paper is as follows: In \sref{sidis}, we summarize the basic setup of SIDIS following the HSO organization of TMD factorization from \cite{Gonzalez-Hernandez:2022ifv}. We also explain the notation to be used throughout the paper. In \sref{pdfsandffs}, we write down the general parametrizations of the TMD pdfs and ffs that we will use for calculations, and in \sref{models} we show how to specialize to specific models of the very small transverse momentum behavior, using Gaussian and spectator-motivated models for illustration. In \sref{largeqtasymptote}, we explain the calculation of the large transverse momentum asymptotic term in the HSO approach. In \sref{matching}, we present sample calculations of the TMD term in SIDIS, with both the Gaussian and spectator inspired models for illustration. After analyzing how the conventional approach to TMD phenomenology leads to the complications discussed above, we show how they are solved in the HSO approach. We end in \sref{conclusion} by discussing future plans for implementing phenomenological treatments in the HSO approach.

\section{Semi-Inclusive DIS}
\label{s.sidis}

\begin{figure}
    \centering
    \includegraphics[width=0.6\textwidth]{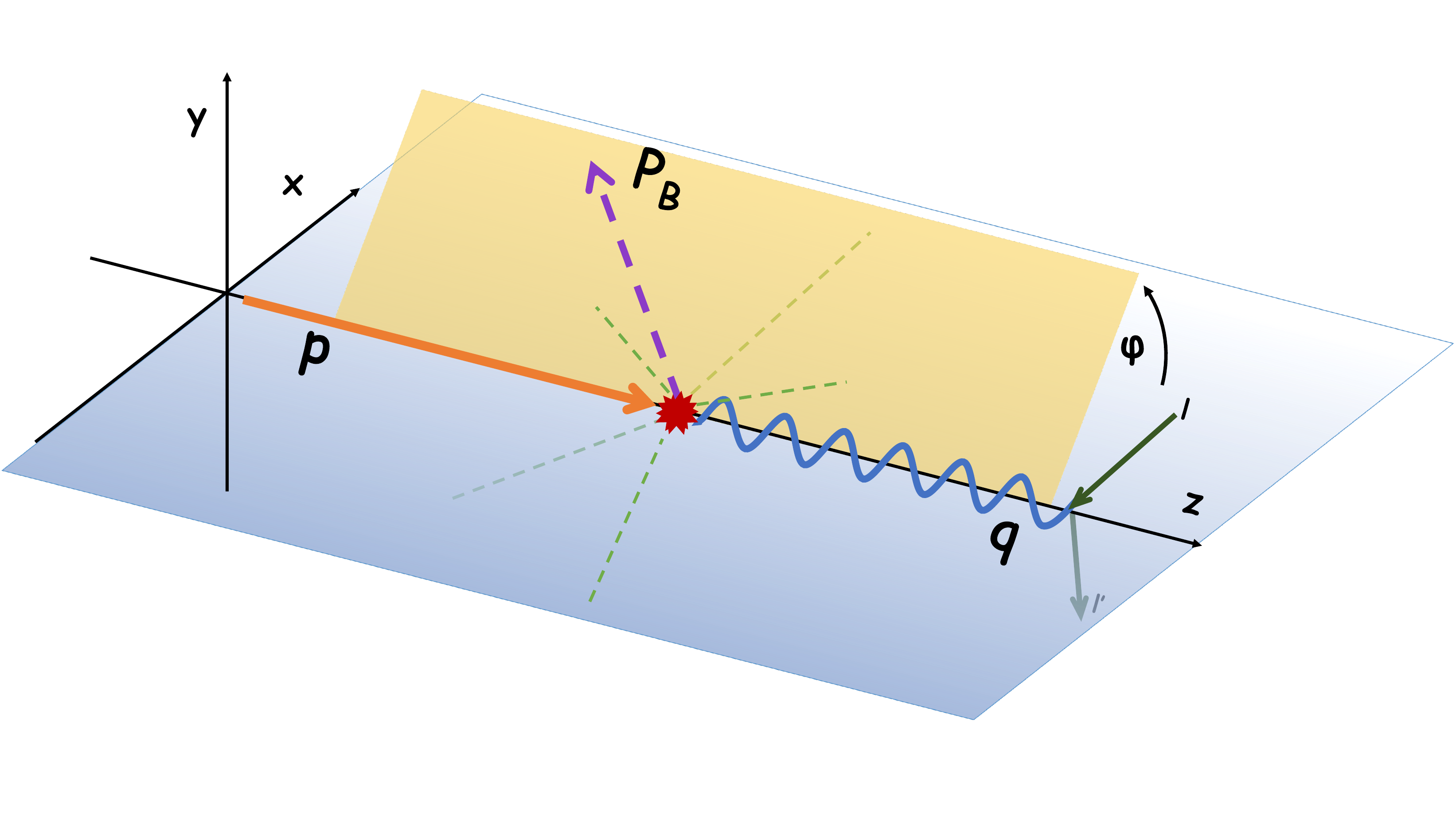}
    \caption{Schematic of a SIDIS event as observed in the Breit frame. The dashed green lines represent the unobserved particles after the collision.}
    \label{fig:SIDIS_Breit_frame}
\end{figure}

We will adopt standard conventions for expressing   SIDIS cross sections in the current fragmentation region, and our labels for the kinematical variables are mostly consistent with those of \cite{Boglione:2019nwk}. 
A lepton with momentum $l$ scatters off a hadron target with momentum $p$, and the momentum of the recoiling lepton is $l'$. The final state contains a measured 
hadron with momentum $P_{\rm B}$ and is inclusive in all other final states $X$:  
\begin{equation}
    l + p\rightarrow l' + P_{\rm B} + X
\end{equation}
Throughout this paper, we will use the usual Lorentz invariant kinematical variables,
\begin{align}
q^2& = -Q^2 \, , &x_\text{bj}& = \frac{Q^2}{2 p \cdot q}\, , 
&z_\text{h}& = \frac{P_B \cdot p}{p \cdot q}\, ,
\end{align}
where $q\equiv (l - l') $ is the momentum of the exchanged photon. Except where specified, we will work in the Breit frame, with the proton moving in the plus light-cone direction {(see figure \ref{fig:SIDIS_Breit_frame})}.
We will drop all power-suppressed target and final state kinematical mass corrections so that Breit frame momentum fractions are 
\begin{align}
    &\xn\equiv -\frac{p^+}{q^+} \approx x_\text{bj},\, . 
\end{align}
where the ``$\approx$'' is a reminder that this identification only holds up to target power suppressed target mass corrections.

For characterizing regions of transverse momentum, we will use the variable 
\begin{equation}
\T{q}{} \equiv - \frac{\T{P}{B}}{\zn} \, .
\end{equation}
Here, $\T{q}{}$ is the transverse momentum of the virtual photon in a frame, which we call the ``hadron frame,'' where the target and final state hadrons are exactly back-to-back, and
\begin{equation}
\zn \equiv \frac{P^-_{\rm B}}{q^-} \, .
\end{equation}
More details concerning the basic kinematical setup that we use may be found in~\cite{Boglione:2019nwk}. In this paper, we will work in a strictly leading power approach, where $x_\text{bj} \approx \xn$ and $z_\text{N} \approx z_\text{h}$. To simplify notation, therefore, we will drop the subscripts on $x$ and $z$ from here forward.

Describing the cross section accurately over the full range of $\Tsc{q}{}$ requires that one merge the treatment tailored to the $\Tsc{q}{}/Q \ll 1$ region (the TMD term) with the collinear factorization treatment appropriate to the $\Tsc{q}{} \approx Q$ region. Both calculations must agree approximately in the intermediate $\Lambda_{\rm QCD} \ll \Tsc{q}{} \ll Q$ region. It is the treatment of the $\Tsc{q}{} \ll Q$ region that involves TMD pdfs and ffs, and it is this contribution that we will focus on in this paper. In the small $\Tsc{q}{}$ limit, and neglecting kinematical hadron mass corrections, $\zn \approx \zh$.

The usual TMD-factorization expression for the hadronic tensor is
\begin{align}
&{} W^{\mu \nu}(\xbj,Q,\zh,\T{P}{B}) \no
&{}= \sum_{j} H_{j}^{\mu \nu} \int \diff{^2 \T{k}{1}}{} \diff{^2 \T{k}{2}}{} f_{j/p}(\xbj,\T{k}{1};\mu_Q,Q^2) D_{h/j}(\zh, \zh\T{k}{2};\mu_Q,Q^2) \delta^{(2)} (\T{q}{} + \T{k}{1} - \T{k}{2}) \no
&{}=\sum_{j} H_{j}^{\mu \nu} \int \frac{\diff[2]{\T{b}{}}}{(2 \pi)^2}
    ~ e^{-i\T{q}{}\cdot \T{b}{} }
    ~ \tilde{f}_{j/p}(\xbj,\T{b}{};\mu_Q,Q^2) 
    ~ \tilde{D}_{h/j}(\zh,\T{b}{};\mu_Q,Q^2) \no
&{}=\sum_{j} H_{j}^{\mu \nu} \left[ f_{j/p}, D_{h/j} \right] \, ,
\label{e.hadrotens}
\end{align}
where the sum is over all quark and antiquark flavors, and each line is a different way that SIDIS routinely gets presented in the literature. The functions $f_{j/p}(\xbj,\T{k}{1};\mu_Q,Q^2)$ and $D_{h/j}(\zh, \zh\T{k}{2};\mu_Q,Q^2)$ are the TMD pdfs and ffs respectively, with their usual operator definitions \cite{Collins:2011qcdbook}. Within the approximations that define TMD factorization in the current region, the longitudinal momentum fractions of the incoming and struck partons are fixed to $\xbj$ and $\zh$. The momentum variables $\T{k}{1}$ and $\T{k}{2}$ are the transverse momenta of the struck and final state partons in the \emph{hadron} frame, and we have fixed the auxiliary renormalization and light-cone scales $\mu$ and $\sqrt{\zeta}$ in \eref{cutoff_def} equal to $\mu_Q$ and $Q$ respectively. (Ultimately, we will set $\mu_Q = Q$, but for organizational purposes we will keep the symbols separate for now.) $H_j^{\mu \nu}$ is a known hard coefficient. 
In transverse coordinate space, the TMD pdfs and ffs are
\begin{align}
\tilde{f}_{j/p}(\xbj,\T{b}{};\mu,\zeta) &{}= \int \diff{^2 \T{k}{1}}{} 
e^{-i \T{k}{1} \cdot \T{b}{} } f_{j/p}(\xbj,\T{k}{1};\mu,\zeta) \, , \no
\tilde{D}_{q/j}(\zh,\T{b}{};\mu,\zeta) &{}= \int \diff{^2 \T{k}{2}}{} 
e^{i \T{k}{2} \cdot \T{b}{} } D_{q/j}(\zh,\zh \T{k}{2};\mu,\zeta) \, ,
\end{align}
and we have used these in the transverse coordinate space representation of $W^{\mu \nu}$ on the third line of \eref{hadrotens}, which is the standard form for implementing evolution.
On the last line of \eref{hadrotens}, we have used the common bracket notation for abbreviating the transverse convolution integrals.
The hard factor is
\begin{equation}
H_{j}^{\mu \nu} = \frac{\zh}{2} {\rm Tr} [\gamma^\nu \gamma^+ \gamma^\mu \gamma^-] |H|^2_{j} \, 
\end{equation}
where the last factor (see, for instance \cite{Collins:2017oxh}) reads

\begin{equation}
    |H|^2_{j} =e_j^2\left\{ 1 
    +\frac{C_F \alpha_s(\mu)}{4\pi}
    \left[
    -16+\frac{\pi^2}{3}+6\ln\left(\frac{Q^2}{\mu^2}\right)-2\ln^2\left(\frac{Q^2}{\mu^2}\right)
    \right]
    +\order{\alpha_s^2(\mu)}    \, \right\}\, . \label{e.hardpart}
\end{equation}
Projection tensors applied to \eref{hadrotens} give the usual unpolarized quark structure functions of SIDIS,
\begin{align}
F_{1,2}\parz{\xbj,Q,\zh,\T{P}{B}} = {\rm P}_{1,2}^{\mu\nu}\, W_{\mu\nu}(\pp,q,\zh,\T{P}{B}),
\end{align}
where, still dropping kinematical hadron mass corrections, 
\begin{subequations}
\label{e.F12proj}
\begin{align}
{\rm P}_1^{\mu\nu}
&= - \frac{1}{2}
   \left[ g^{\mu\nu}
	- 4\xbj^2 \frac{\pp^\mu \pp^\nu}{Q^2}
   \right] \, ,
\label{e.P1}\\
{\rm P}_2^{\mu\nu}
&= - \xbj
   \left[ g^{\mu\nu}
	- 12 \xbj^2
	  \frac{\pp^\mu \pp^\nu}{Q^2}
   \right]\, ,
\label{e.P2}
\end{align}
\end{subequations}
and
\begin{align}
&{}{\rm P}_1^{\mu\nu} H_{j,\mu \nu} = H_1 = 2 \zh |H|^2_{j} \, , \no
&{}{\rm P}_2^{\mu\nu} H_{j,\mu \nu} = H_2 = 4 \zh \xbj |H|^2_{j} \, .
\end{align}

Reference~\cite{Gonzalez-Hernandez:2022ifv} substantially reorganized the more standard ways of expressing the TMD factorization expression for $W^{\mu \nu}$, as summarized by the sequence of steps in Sec.~VI of that paper. Doing so required a high degree of specificity about exactly which versions of pdfs and ffs and their parametrizations were being discussed in a given context, and this led us to introduce a rather elaborate system of notation. For conciseness, we will drop most of that notation in this paper and instead indicate in the text which version of a symbol is being used whenever such distinctions become necessary. When we calculate \eref{hadrotens}, we will mostly be interested in using the final underlined $\tilde{f}_{j/p}$, $\tilde{D}_{h/j}$ in Eq.~(61) of~\cite{Gonzalez-Hernandez:2022ifv}, although for the input scale calculations in this paper the difference between the underlined and ``input'' distributions is negligible. Any perturbatively calculable quantities will be  maintained through order $\alpha_s$, so results are all $\order{\alpha_s}$. Any collinear pdfs or ffs should be assumed to be defined in the $\msbar$ renormalization scheme unless otherwise specified. Power suppressed errors will be expressed as $\order{m^2/Q^2}$ where $m$ symbolizes any small mass scale like $\Lambda_\text{QCD}$ or a hadron mass. 

To implement evolution, we rewrite \eref{hadrotens} in a form where each TMD function is expressed in terms of evolution from an input scale $Q_0$. 
Thus, we use (the SIDIS version of) Eq.~(65) from \cite{Gonzalez-Hernandez:2022ifv},
\begin{align} 
W^{\mu \nu}(\xbj,Q,\zh,\T{P}{B})
&{}=
    \sum_{j} H_{j}^{\mu \nu} \int \frac{\diff[2]{\T{b}{}}}{(2 \pi)^2}
    ~ e^{-i\T{q}{}\cdot \T{b}{} }
    ~ \tilde{f}_{j/p}(\xbj,\T{b}{};\mu_{Q_0},Q_0^2) 
    ~ \tilde{D}_{h/j}(\zh,\T{b}{};\mu_{Q_0},Q_0^2) \nonumber\\&
  \,\times
  \exp\left\{  
        \tilde{K}(\Tsc{b}{};\mu_{Q_0}) \ln \parz{\frac{ Q^2 }{ Q_0^2}}
           +\int_{\mu_{Q_0}}^{\mu_Q}  \frac{ \diff{\mu'} }{ \mu' }
           \biggl[ 2 \gamma(\alpha_s(\mu'); 1) 
                 - \ln\frac{Q^2}{ {\mu'}^2 } \gamma_K(\alpha_s(\mu'))
           \biggr]
  \right\} \, .
\label{e.Wtermev0_finalversion}
\end{align}
$Q_0$ should be understood to be the lowest value of $Q$ for which factorization techniques are considered reasonable, which in practice is usually between around $1$~GeV and $4$~GeV for SIDIS. An important observation underlying the HSO approach of Ref.~\cite{Gonzalez-Hernandez:2022ifv} is that individual correlation functions, $f_{j/p}(\xbj,\T{k}{};\mu_{Q_0},Q_0^2)$ or $\tilde{D}_{h/j}(\zh,\T{b}{};\mu_{Q_0},Q_0^2)$, have unambiguous transverse momentum dependence for all $\Tsc{k}{}$, including all $\Tsc{k}{} > Q_0$, which follows from their operator definitions. Once these input functions have been  determined, evolving them to larger $Q$ is only a matter of substituting them into \eref{Wtermev0_finalversion} (after transforming into coordinate space). This can be used to simplify the organization of phenomenological implementations because one may focus attention on the nonperturbative momentum space treatment of hadron structure at $Q$ near the initial input scale $Q_0$. The only input that is then necessary to obtain the TMDs at any other higher scale is the evolution kernel.

In this paper, we will be mostly interested in the behavior of the input TMD pdfs and ffs, in which case the evolution factor does not enter. In places where we do need the evolution factor, we will use the same parametrization for the CS kernel 
from Sec.~VII-A from Ref.~\cite{Gonzalez-Hernandez:2022ifv} 
since it reproduces the correct $\order{\alpha_s}$ perturbative behavior while also capturing minimal basic expectations for the nonperturbative behavior.
Thus, the input scale parametrization of the kernel that we will use is
\begin{align}
\inpt{\tilde{K}}(\Tsc{b}{};\mu_{Q_0})
= \frac{2\alpha_s(\mu_{Q_0}) C_F}{\pi} 
\left[ K_0(\Tsc {b}{} m_K) + 
 \ln\parz{\frac{m_K}{\mu_{Q_0}}} \right] \, 
\label{e.K_param}
\end{align}
so the full (underlined, in the notation of Ref.~\cite{Gonzalez-Hernandez:2022ifv}) kernel is
\begin{align}
\tilde{K}(\Tsc{b}{};\mu_{Q_0}) 
&{}= 
\frac{2\alpha_s(\mu_{\overline{Q}_0}) C_F}{\pi} 
\left[ K_0(\Tsc {b}{} m_K) + 
 \ln\parz{\frac{m_K}{\mu_{\overline{Q}_0}}} \right] -
\int_{\mu_{\overline{Q}_0}}^{\mu_{Q_0}} \frac{\diff{\mu'}}{\mu'}  \gamma_K(\alpha_s(\mu')) \, .
\label{e.K_param_final}
\end{align} 
The nonperturbative model parameter in $\tilde{K}(\Tsc{b}{};\mu_{Q_0})$ is $m_K$. The bar on top of $\overline{Q}_0$ and $\mu_{\overline{Q}_0}$ is the symbol introduced in~\cite{Gonzalez-Hernandez:2022ifv} to indicate that this is a scale that is fixed to $Q_0$ at large $\Tsc{b}{}$, but which transitions to $\sim 1/\Tsc{b}{}$ behavior as $\Tsc{b}{} \to 0$. 
 The role of the ``scale transformation function", $\overline{Q}_0$, is analogous to that of $b_*$ in the usual CSS treatment, and its exact choice is, in principle, arbitrary. We will continue to use the choice for $\overline{Q}_0$ from Ref.~\cite{Gonzalez-Hernandez:2022ifv}. We provide the expression in \aref{interp} of this paper.
We remark that it is possible to consider other types of nonperturbative behavior for the CS kernel within the approach of  Ref.~\cite{Gonzalez-Hernandez:2022ifv}, including recent calculations in lattice QCD (see for instance Refs.~\cite{Schlemmer:2021aij, Li:2021wvl,Shanahan:2021tst,LPC:2022ibr}).

\section{TMD parton density \& fragmentation functions}
\label{s.pdfsandffs}

For constructing parametrizations of the quark and antiquark TMD pdfs and ffs, we repeat the steps in Sec.VI of Ref.~\cite{Gonzalez-Hernandez:2022ifv}. We continue to use the additive structure from the examples in Ref.~\cite{Gonzalez-Hernandez:2022ifv} to interpolate between a nonperturbative core and the perturbative tail. The first terms transition into the fixed $\order{\alpha_s(\mu)}$ tail calculation of the TMD at large $\Tsc{k}{}$, while the last term is a non-perturbative ``core'' that describes the peak at very small $\Tsc{k}{}$. 
The core term is further constrained by an integral relation analogous to \eref{cutoff_def}, which determines its overall normalization factor $C_{h/j}$.

Thus, for the input quark ff
\begin{align}
\inptp{D}{h/j}(\zh,\zh \T{k}{};\mu_{Q_0},Q_0^2) &{}=  \frac{1}{2 \pi \zh^2} \frac{1}{\Tscsq{k}{} + m_{D_{h,j}}^2} \left[A^D_{h/j}(\zh;\mu_{Q_0}) + B^D_{h/j}(\zh;\mu_{Q_0}) \ln \frac{Q_0^2 }{\Tscsq{k}{}+m_{D_{h,j}}^2} \right] \no
&{}+\frac{1}{2 \pi \zh^2} \frac{1}{\Tscsq{k}{} + m_{D_{h,g}}^2} A^{D,g}_{h/j}(\zh;\mu_{Q_0})   \no
&{} 
+
C^D_{h/j} \,\np{D}{h/j}(\zh,\zh \T{k}{};Q_0^2) \, , 
\label{e.candidateqff}
\end{align}
where $\np{D}{h/j}(\zh,\zh \T{k}{};Q_0^2)$ is a parametrization of the peak of the TMD ff to be specified later. 
To compactify notation, we have dropped the $(n,d_r)$ superscripts that were used in \cite{Gonzalez-Hernandez:2022ifv}, but we have included a hadron label $h$ and $j \in {u,d,s,c,\dots}$ labels for parton flavors and anti-flavors. $A^D$, $B^D$, and $C^D$ are abbreviations for the following expressions,
\begin{align}
&{} A^D_{h/j}(z;\mu_{Q_0}) \equiv \sum_{j j'} \delta_{j'j} \frac{\alpha_s(\mu_{Q_0})}{\pi}
\left\{ \left[(P_{jj'}\otimes{d}_{h/j'})(z;\mu_{Q_0})\right] \vphantom{\frac{3 C_F}{2} d(z;\mu_{Q_0})} -  \frac{3 C_F}{2} d_{h/j'}(z;\mu_{Q_0})  \right\} \, , \label{e.A_def}  \\
&{} B^D_{h/j}(z;\mu_{Q_0}) \equiv \sum_{j j'} \delta_{j' j} \frac{\alpha_s(\mu_{Q_0}) C_F}{\pi}d_{h/j'}(z;\mu_{Q_0}) \, , \label{e.B_def} \\
&{}A^{D,g}_{h/j}(z;\mu_{Q_0}) \equiv \frac{\alpha_s(\mu_{Q_0})}{\pi} \left[(P_{gj}\otimes{d}_{h/g})(z;\mu_{Q_0})\right] \, , \\
&{}C^D_{h/j} \equiv
\frac{1}{N^D_{h/j}}
\Bigg[
d_{h/j}(z;\mu_{Q_0}) - A^D_{h/j}(z;\mu_{Q_0}) \ln \parz{\frac{\mu_{Q_0}}{m_{D_{h,j}}}} -  B^D_{h/j}(z;\mu_{Q_0}) \ln \parz{\frac{\mu_{Q_0}}{m_{D_{h,j}}}} \ln \parz{\frac{Q_0^2}{\mu_{Q_0} m_{D_{h,j}}} } \, ,  \no
&{} - A^{D,g}_{h/j}(z;\mu_{Q_0}) \ln \parz{\frac{\mu_{Q_0}}{m_{D_{h,g}}}} +\frac{\alpha_s(\mu_{Q_0})}{2 \pi} 
\left\{ \sum_{j j'} \delta_{j' j} [ \mathcal{C}_{\Delta}^{j'/j} \otimes d_{h/j'} ](z;\mu_{Q_0}) + [ \mathcal{C}_{\Delta}^{g/j} \otimes d_{h/g} ](z;\mu_{Q_0}) \right\}
\Bigg]
\, . \label{e.C_def} 
\end{align}
where 
\begin{align}
P_{qq}(z)&{}=P_{\bar{q}\bar{q}}(z)=C_F\left[\frac{1+z^2}{\parz{1-z}_+}+\frac{3}{2}\delta\parz{1-z}\right] \, , \label{e.pdists} \\
P_{gq}(z)&{}= C_F \frac{1 + (1-z)^2}{z} \, , \\
\mathcal{C}_{\Delta}^{q/q}(z) &{}= 2 P_{qq}(z) \ln z  + C_F (1 - z) -C_F\frac{\pi^2}{12}\delta(1-z)  \, ,
\label{e.Delta_trans1} \\
\mathcal{C}_{\Delta}^{g/q}(z) &{}= 2 P_{gq}(z) \ln z + C_F z  \, 
\label{e.Delta_trans2},\\
N^D_{h/j}\equiv{}&
\,2\pi\,z^2\int_{0}^{\infty}d\Tsc{k}{}\Tsc{k}{}\,\np{D}{h/j}(z,z \T{k}{};Q_0^2) \, .
\label{e.Dnorm}
\end{align}
For the TMD pdfs, the expressions are similar,
\begin{align}
\inptp{f}{i/p}(\xbj,\T{k}{};\mu_{Q_0},Q_0^2) &{}= 
\frac{1}{2 \pi} \frac{1}{\Tscsq{k}{} + m_{f_{i,p}}^2} \left[A^f_{i/p}(\xbj;\mu_{Q_0}) + B^f_{i/p}(\xbj;\mu_{Q_0}) \ln \frac{Q_0^2 }{\Tscsq{k}{}+m_{f_{i,p}}^2} \right] \no
&{}+  \frac{1}{2 \pi} \frac{1}{\Tscsq{k}{} + m_{f_{g,p}}^2} A_{i/p}^{f,g}(\xbj;\mu_{Q_0}) \no
&{} 
+
C^f_{i/p} \,\np{f}{i/p}(\xbj,\T{k}{};Q_0^2) \, ,
\label{e.candidateqpdf}
\end{align}
with the corresponding abbreviations
\begin{align}
&{} A^f_{i/p}(x;\mu_{Q_0}) \equiv \sum_{ii'} \delta_{i'i} \frac{\alpha_s(\mu_{Q_0})}{\pi}
\left\{ \left[(P_{i'i}\otimes{f}_{i'/p})(x;\mu_{Q_0})\right] \vphantom{\frac{3 C_F}{2} d(z;\mu_{Q_0})} -  \frac{3 C_F}{2} f_{i'/p}(x;\mu_{Q_0})  \right\} \, , \label{e.A_def_pdf}  \\
&{} B^f_{i/p}(x;\mu_{Q_0}) \equiv \sum_{i'i} \delta_{i'i} \frac{\alpha_s(\mu_{Q_0}) C_F}{\pi}f_{i'/p}(x;\mu_{Q_0})  \, , \label{e.B_def_pdf} \\
&{}A^{f,g}_{i/p}(x;\mu_{Q_0}) \equiv \frac{\alpha_s(\mu_{Q_0})}{\pi} \left[(P_{ig}\otimes{f}_{g/p})(x;\mu_{Q_0})\right] \, , \\
&{}C^f_{i/p} \equiv
\frac{1}{N^f_{i/p}}
\Bigg[
f_{i/p}(x;\mu_{Q_0}) - A^f_{i/p}(x;\mu_{Q_0}) \ln \parz{\frac{\mu_{Q_0}}{m_{f_{i,p}}}} -  B^f_{i/p}(x;\mu_{Q_0}) \ln \parz{\frac{\mu_{Q_0}}{m_{f_{i,p}}}} \ln \parz{\frac{Q_0^2}{\mu_{Q_0} m_{f_{i,p}}} } \, ,  \no
&{} - A^{f,g}_{i/p}(x;\mu_{Q_0}) \ln \parz{\frac{\mu_{Q_0}}{m_{f_{g,p}}}} +\frac{\alpha_s(\mu_{Q_0})}{2 \pi} 
\left\{ \sum_{ii'} \delta_{i'i} [ \mathcal{C}_{\Delta}^{i/i'} \otimes f_{i'/p} ](x;\mu_{Q_0}) + [ \mathcal{C}_{\Delta}^{i/g} \otimes f_{g/p} ](x;\mu_{Q_0}) \right\}
\Bigg]
\, . \label{e.C_def_pdf} 
\end{align}
where 
\begin{align}
P_{ig}(x)&{}= T_F \left[x^2 + (1-x)^2 \right] \, , \\
\mathcal{C}_{\Delta}^{i/i}(x) &{}=   C_F (1 - x) -C_F\frac{\pi^2}{12}\delta(1-x) \, ,
\label{e.Delta_trans1pdf} \\
\mathcal{C}_{\Delta}^{g/p}(x) &{}=  2 T_F x (1-x)  \, ,
\label{e.Delta_trans2pdf} \\
N^f_{i/p}\equiv{}&
\,2\pi\,\int_{0}^{\infty}d\Tsc{k}{}\Tsc{k}{}\,\np{f}{i/p}(x, \T{k}{};Q_0^2)
\label{e.Fnorm}
\end{align}
In \eref{candidateqpdf}, $\np{f}{i/p}(x, \T{k}{};Q_0^2)$ parametrizes the core peak of the TMD pdf.
(We remind the reader that it is to be understood that all explicit perturbative parts in this paper are calculated to lowest order in $\alpha_s$.) 

To extend the TMD pdf and ff parametrizations above to account for the $\Tsc{b}{} \ll 1/Q_0$ region, we transform to transverse coordinate space and use Eq.~(92) of \cite{Gonzalez-Hernandez:2022ifv} and its analog for the  TMD pdf,
\begin{align}
\tilde{D}_{h/j}(z,\T{b}{};\mu_{Q_0},Q_0^2) = \inptp{\tilde{D}}{h/j}(z,\T{b}{};\mu_{\overline{Q}_0},\overline{Q}_0^2) E(\overline{Q}_0/Q_0,\Tsc{b}{}) \, . \label{e.evolvedd4}
\end{align} 
\begin{align}
\tilde{f}_{i/p}(x,\T{b}{};\mu_{Q_0},Q_0^2) = \inptp{\tilde{f}}{i/p}(x,\T{b}{};\mu_{\overline{Q}_0},\overline{Q}_0^2) E(\overline{Q}_0/Q_0,\Tsc{b}{})\, . \label{e.evolvedd4pdf}
\end{align} 
with an evolution factor
\begin{equation}
E(\overline{Q}_0/Q_0,\Tsc{b}{}) \equiv \exp \left\{
\int_{\mu_{\overline{Q}_0}}^{\mu_{Q_0}} \frac{d \mu^\prime}{\mu^\prime} \left[\gamma(\alpha_s(\mu^\prime);1) 
- \ln \frac{Q_0}{\mu^\prime} \gamma_K(\alpha_s(\mu^\prime))
  \right] +\ln \frac{Q_0}{\overline{Q}_0} \inpt{\tilde{K}}(\Tsc{b}{};\mu_{\overline{Q}_0}) \right\}
  \label{e.ev_factor} \, .
\end{equation}
Once the numerical values of parameters in $\tilde{D}_{h/j}(z,\T{b}{};\mu_{Q_0},Q_0^2)$ and $\tilde{f}_{i/p}(x,\T{b}{};\mu_{Q_0},Q_0^2)$ are determined and fixed as above, the TMD term at any other larger scale $Q$ is found straightforwardly by substituting these into \eref{Wtermev0_finalversion}. 

The scale $\overline{Q}_0$ is designed to be approximately $Q_0$ for $Q \approx Q_0$, where the only important range of $\Tsc{b}{}$ is $\Tsc{b}{} \gtrsim 1/Q_0$, and the left and right sides of \erefs{evolvedd4}{evolvedd4pdf} are nearly equal. For large $Q$ ($Q \gg Q_0$), the UV $\Tsc{b}{} \ll 1/Q_0$ region starts to become important and cannot be ignored. There, $\overline{Q}_0$ smoothly transitions into a $\sim 1/\Tsc{b}{}$ behavior such that RG improvement is implemented in the $\T{b}{} \to \T{0}{}$ limit. The left sides of \erefs{evolvedd4}{evolvedd4pdf} are the parametrizations that we labeled with underlines in Eq.(60) of Ref.~\cite{Gonzalez-Hernandez:2022ifv}, while the ``input'' functions on the left sides are to be used for phenomenological fitting for $Q \approx Q_0$. By construction, the left and right sides of \erefs{evolvedd4}{evolvedd4pdf}, as well $Q_0$ and $\overline{Q}_0$,  differ negligibly in the range of $\Tsc{b}{}$ relevant to $Q \approx Q_0$ phenomenology -- recall the discussion in Sec.~V of \cite{Gonzalez-Hernandez:2022ifv}. 

For the examples implementations we will perform in \sref{hso}, we will use the approximation 
\begin{equation}
\label{e.Eapprox}
E(\overline{Q}_0/Q_0,\Tsc{b}{}) \approx 1 \, ,
\end{equation}
and set $\overline{Q}_0 \to Q_0$, 
since for this paper our main focus is on the $Q \approx Q_0$ region
and the construction of satisfactory parametrizations for $\inptp{\tilde{D}}{h/j}(z,\T{b}{};\mu_{Q_0},Q_0^2)$ and $\inptp{\tilde{f}}{i/p}(x,\T{b}{};\mu_{Q_0},Q_0^2)$. At the end of \sref{hso}, 
we will restore the $\overline{Q}_0$ treatment and confirm that its effect is negligible at $Q \approx Q_0$.

It can be seen by inspection that the input parametrizations defined in \eref{candidateqff} and \eref{candidateqpdf} are constrained to match the perturbative large-$\Tsc{k}{}$ collinear factorization approximations for the TMD pdfs and ffs,
\begin{align}
\inptp{D}{h/j}^\text{pert}(z,z \T{k}{};\mu_{Q_0},Q_0^2) &{}=  \frac{1}{2 \pi z^2} \frac{1}{\Tscsq{k}{}} \left[A^D_{h/j}(z;\mu_{Q_0}) + B^D_{h/j}(z;\mu_{Q_0}) \ln \frac{Q_0^2 }{\Tscsq{k}{}} \right] +\frac{1}{2 \pi z^2} \frac{1}{\Tscsq{k}{}} A^{D,g}_{h/j}(z;\mu_{Q_0}) \, , \label{e.candidateqff_pert} \\
\inptp{f}{i/p}^\text{pert}(x,\T{k}{};\mu_{Q_0},Q_0^2) &{}= 
\frac{1}{2 \pi} \frac{1}{\Tscsq{k}{}} \left[A^f_{i/p}(x;\mu_{Q_0}) + B^f_{i/p}(x;\mu_{Q_0}) \ln \frac{Q_0^2 }{\Tscsq{k}{}} \right] + \frac{1}{2 \pi} \frac{1}{\Tscsq{k}{}} A_{i/p}^{f,g}(x;\mu_{Q_0})  \, , \label{e.candidateqpdf_pert}
\end{align}
which are good approximations to the true TMD correlation functions when $\Tsc{k}{} \approx Q_0$ and $Q_0 \gg m$. Equations~\eqref{e.candidateqff_pert} and \eqref{e.candidateqpdf_pert} are calculable entirely within leading power collinear factorization. The same expressions apply at any value of $Q$, but for this paper we are especially interested in $Q$ near the input scale. 

\section{Gaussian versus scalar diquark models}
\label{s.models}

The model parametrizations of the last section are still quite general. The only choices that have been made so far are to use an additive structure to interpolate to the order-$\alpha_s$ perturbative tail at $\Tsc{k}{} \approx Q_0$ and the choice of the parametrization of the CS kernel in \eref{K_param_final}. Further assumptions are necessary before these parametrizations can become useful.

Most of the effort in nonperturbative modeling enters in the choices for the functional forms for $\np{D}{h/j}(\zh,\zh \T{k}{};Q_0^2)$ and $\np{f}{i/p}(\xbj,\T{k}{};Q_0^2)$ that describe the very small $\Tsc{k}{} \approx \Tsc{0}{}$ behavior.  However, many approaches to modeling or parametrizing this region of nonperturbative TMDs already exist \cite{Kotzinian:1994dv,Gamberg:2003ey,Bacchetta:2008af,Kang:2010hg,Guerrero:2020hom,Pasquini:2008ax,Pasquini:2011tk,Bacchetta:2017vzh,Pasquini:2014ppa,Hu:2022ctr,Sakai:1979rf,Sakai:1979ez,Yuan:2003wk,Avakian:2010br,Signal:2021aum,Matevosyan:2011vj,Matevosyan:2012ga,Noguera:2015iia,Shi:2018zqd,Broniowski:2017wbr,Bastami:2020asv,Bastami:2020rxn}, and one may defer to them at this stage in the parametrization construction. The only way these previously existing models need to be modified is by including the interpolation to the order $\alpha_s$ large-$\Tsc{k}{}$ behavior, and by imposing integral relations analogous to \eref{cutoff_def}. 
All that remains is to adjust $\np{D}{h/j}(\zh,\zh \T{k}{};Q_0^2)$ and $\np{f}{i/p}(\xbj,\T{k}{};Q_0^2)$ so as to recover (at least approximately) existing model parametrizations in the $\Tsc{k}{} \approx 0$ region. The parameters $m_{D_{j,h}},m_{D_{g,h}},m_{f_{i,p}},m_{f_{g,p}}$ control the transition between the $\Tsc{k}{}$ model and the large $\Tsc{k}{}$ perturbative tail. 

For the purposes of this article, we will focus on two of the most commonly used models in phenomenology that are simple to implement. The first is the Gaussian model of TMDs (see, for example, Refs.\cite{Schweitzer:2010tt,Anselmino:2013vqa,Anselmino:2013lza}), which is often found to successfully describe data at lower $Q$. It prescribes the functions forms
\begin{align}
\label{e.npmodels}
\np{f}{i/p}^{\text{Gauss}}(\xbj,\T{k}{};Q_0^2)={}&
\frac{e^{-\Tscsq{k}{}/M_\text{F}^2}}{\pi  M_\text{F}^2}\,,
\qquad
\np{D}{h/j}^{\text{Gauss}}(\zh,\zh \T{k}{};Q_0^2)={}
\frac{e^{-\zh^2\,\Tscsq{k}{}/M_\text{D}^2}}{\pi  M_\text{D}^2}\, .
\end{align}
The second model that we will consider is 
inspired by the popular spectator diquark model~\cite{Bacchetta:2007wc,Bacchetta:2008af}.
For it, we adopt the functional forms
\begin{align}
\label{e.spectatorpdf}
\np{f}{i/p}^{\text{Spect}}(\xbj,\T{k}{};Q_0^2)={}&
\frac{6 \mospec^6}{\pi  \left(2 \mspec^2+\mospec^2\right)}
\,
\frac{\mspec^2+\Tscsq{k}{}}{\left(\mospec^2+\Tscsq{k}{}\right){}^4}
\,.
\\
\label{e.spectatorff}
\np{D}{h/j}^{\text{Spect}}(\zh,\zh \T{k}{};Q_0^2)={}&
\frac{2 \mospecff^4}{\pi  \left(\mspecff^2+\mospecff^2\right)}
\,
\frac{\mspecff^2+\Tscsq{k}{} \zh^2}{\left(\mospecff^2+\Tscsq{k}{} \zh^2\right){}^3}\,,
\end{align}
The overall factors in \erefs{npmodels}{spectatorff} are chosen so that $N^D_{h/j} = N^f_{i/p} = 1$ in both models (recall \eref{Dnorm} and \eref{Fnorm}). 

In later sections, it will often be convenient to work with collinear pdfs and ffs defined as the cutoff transverse momentum integrals of TMD pdfs and ffs. Hence, we define
\begin{align}
f^c_{i/p}(x;\mu_Q) &{}\equiv 2 \pi \int_0^{\mu_Q} \diff{\Tsc{k}{}}{} \Tsc{k}{} f_{i/p}(x,\T{k}{};\mu_Q,Q^2) \, , \label{e.fc_deff} \\
d^c_{h/j}(z;\mu_Q) &{}\equiv 2 \pi z^2 \int_0^{\mu_Q} \diff{\Tsc{k}{}}{} \Tsc{k}{} D_{h/j}(z,z \T{k}{};\mu_Q,Q^2) \, , \label{e.dc_deff}
\end{align}
where the $c$ superscript stands for ``cutoff.'' The cutoff definitions could be defined more generally with an upper limit $\mu_f$ different from $\mu_{Q}$, but we will keep these scales equal for the present paper. The cutoff and $\msbar$-renormalized definitions are equal up to a scheme change and $m^2/\mu_Q^2$-suppressed corrections. 

With our parametrizations of TMD pdfs and ffs in the previous section, the integrals are 
\begin{align}
\inptp{f}{i/p}^c(x;\mu_{Q_0}) =&{} 2 \pi \int_0^{\mu_{Q_0}} \diff{\Tsc{k}{}}{} \Tsc{k}{} \inptp{f}{i/p}(x,\T{k}{};\mu_{Q_0},Q_0^2) = \no
&{} 
C^f_{i/p}\, \np{f}{i/p}^c(x;\mu_{Q_0})  + 
\frac{1}{2}
A^{f,g}_{i/p}
(x;\mu_{Q_0})
\ln \parz{1 + \frac{\mu_{Q_0}^2}{m_{f_{g,p}}^2}} \no 
&{} + 
\frac{1}{2} A^f_{i/p}(x;\mu_{Q_0}) \ln \parz{1 + \frac{\mu_{Q_0}^2}{m_{f_{i,p}}^2}}
+
\frac{1}{4} B^f_{i/p}(x;\mu_{Q_0}) \left[\ln^2\parz{\frac{m_{f_{i,p}}^2}{Q_0^2}} - \ln^2{\parz{\frac{\mu_{Q_0}^2+ m_{f_{i,p}}^2}{Q_0^2}}} \right] \no
&= f_{i/p}(x;\mu_{Q_0}) + \order{\alpha_s(\mu_0)^2,\frac{m^2}{Q_0^2}} \, , \label{e.cutpdf_param}
\end{align}
and
\begin{align}
\inptp{d}{h/j}^c(z;\mu_{Q_0}) =&{} 2 \pi z^2 \int_0^{\mu_{Q_0}} \diff{\Tsc{k}{}}{} \Tsc{k}{} \inptp{D}{h/j}(z,z \T{k}{};\mu_{Q_0},Q_0^2) = \no 
&{}
C^D_{h/j} \, \np{d}{h/j}^c(z;\mu_{Q_0}) +
\frac{1}{2} 
A^{D,g}_{h/j}(z;\mu_{Q_0})
\ln \parz{1 + \frac{\mu_{Q_0}^2}{m_{D_{h,g}}^2}} \no 
&{} + 
\frac{1}{2} 
A^D_{h/j}(z;\mu_{Q_0}) \ln \parz{1 + \frac{\mu_{Q_0}^2}{m_{D_{h,j}}^2}}
+ 
\frac{1}{4} B^D_{h/j}(z;\mu_{Q_0}) \left[\ln^2\parz{\frac{m_{D_{h,j}}^2}{Q_0^2}} - \ln^2{\parz{\frac{\mu_{Q_0}^2+ m_{D_{h,j}}^2}{Q_0^2}}} \right] \,  \no
&= d_{h/j}(z;\mu_{Q_0}) + \order{\alpha_s(\mu_0)^2,\frac{m^2}{Q_0^2}} \, ,
\label{e.cutff_param}
\end{align}
with
\begin{align}
\label{e.npmodels_cutoff}
\np{f}{i/p}^{c,\text{Gauss}}(\xbj;\mu_{Q_0},Q_0^2)={}&
1-e^{-\mu_{Q_0}^2/M_\text{F}^2}\,,
\qquad
\np{d}{h/j}^{c,\text{Gauss}}(\zh;\mu_{Q_0},Q_0^2)={}
1-e^{-\zh^2\,\mu_{Q_0}^2/M_\text{D}^2}\,,
\end{align}
in the case of the Gaussian model, and 
\begin{align}
\label{e.spectatorpdf_cutoff}
\np{f}{i/p}^{c,\text{Spect}}(\xbj;\mu_{Q_0},Q_0^2)={}&
1-\frac{\mospec^6 \left(2 \mspec^2+\mospec^2+3 \mu_{Q_0} ^2\right)}{\left(2
   \mspec^2+\mospec^2\right) \left(\mospec^2+\mu_{Q_0} ^2\right){}^3}\,,
\\
\label{e.spectatorff_cutoff}
\np{d}{h/j}^{c,\text{Spect}}(\zh;\mu_{Q_0},Q_0^2)={}&
1-\frac{\mospecff^4 \left(\mspecff^2+\mospecff^2+2 \mu_{Q_0} ^2
   z^2\right)}{\left(\mspecff^2+\mospecff^2\right) \left(\mospecff^2+\mu_{Q_0} ^2 z^2\right){}^2}
\,.
\end{align}
in the case of the spectator model. Note that \erefs{npmodels_cutoff}{spectatorff_cutoff} are all $1$ up to (at most) $m^2/\mu_{Q_0}^2$-suppressed errors. 

The expressions in \erefs{cutpdf_param}{cutff_param} follow directly by substituting 
\eref{candidateqff} and \eref{candidateqpdf} into 
\erefs{fc_deff}{dc_deff}.
By substituting the expressions 
in \eref{C_def} and \eref{C_def_pdf} for $C^D_{h/j}$ and $C^f_{i/p}$, it is straightforward to verify that 
the collinear pdfs and ffs of \eref{cutpdf_param}  and 
\eref{cutff_param} 
are equal to the standard \msbar $\,\,f(x;\mu_{Q_0})$ and $d(z;\mu_{Q_0})$ respectively in the limit that $\order{m^2/Q_0^2}$ and $\order{\alpha_s(\mu_{Q_0})^2}$ errors are negligible.

To further simplify later numerical examples and reduce the number of free parameters,
in the spectator-like models we will fix $\mospec=\mospecff/z=0.2\,\text{GeV}$. 
We will also assume that model masses 
have no parton flavor dependence, and that the $M_{F,D}$ of the core distributions is the same as the $m_{f,D}$ in the tail terms. That is, 
for both models 
we will assume for now
\begin{align}
&{}m_{f_{i,p}} = m_{f_{g,p}} = M_F \, , \label{e.spm1} \\
&{}m_{D_{h,j}} = m_{D_{h,g}} = M_D \, . \label{e.spm2}
\end{align}
In general, the parameters in \erefs{spm1}{spm2} could have different numerical values 
in the Gaussian and the spectator-like models, 
but we will keep the same labels in both to simplify notation.

It should be emphasized that nothing in the setup of \sref{pdfsandffs} relies on the use of any \emph{particular} nonperturbative model. Indeed, one of the motivating advantages of the HSO approach is that the momentum space nonperturbative model of the $\Tsc{k}{} \approx 0$ region becomes easily interchangeable, as demonstrated by our switching between the Gaussian and spectator diquark models above. 

\section{The large transverse momentum asymptote}
\label{s.largeqtasymptote}

In this section, we will enumerate the steps for extracting the large-$\Tsc{q}{}$ asymptote of \eref{Wtermev0_finalversion}. These steps will be particularly relevant to phenomenological treatments of the $Q \approx Q_0$ region. Our path here differs from that of more standard presentations in that we start with the \emph{small} transverse momentum part of the TMD term and extract the large $\Tsc{q}{} \approx Q$ behavior, in contrast to the more usual steps that start with large-$\Tsc{q}{}$ calculations of the cross section in collinear perturbation theory and extract the $\Tsc{q}{} \to 0$ asymptote. Both approaches must give the same result up to $\order{m^2/Q^2}$ and $\order{\alpha_s(Q)^{n+1}}$ corrections. 

In all steps below, we will assume we are analyzing the TMD term in a regime where $\Tsc{q}{}$ is comparable to $Q$ and $Q$ approaches infinity.  To be specific, we take $\Tsc{q}{} = \eta Q$ where $\eta$ is a fixed, order unity constant and we let $m^2/Q^2 \to 0$. 
It will be convenient to first express the transverse momentum convolution integral on the second line of \eref{hadrotens} in the following way, 
\begin{equation}
\label{e.rewrite}
\left[ f, D \right] = \int \diff{^2 \T{k}{}}{} f(\xbj,\T{k}{}-\T{q}{}/2;\mu_Q;Q^2) D(\zh,\zh \parz{\T{k}{}+\T{q}{}/2};\mu_Q;Q^2) \, , 
\end{equation}
where flavor subscripts are dropped. If we first consider the region of the integrand where 
\begin{equation}
\T{k}{} = \T{q}{}/2 + \order{m} \, , 
\end{equation}
then 
\begin{align}
&{} f(\xbj,\T{k}{}-\T{q}{}/2;\mu_Q;Q^2) D(\zh,\zh\parz{\T{k}{}+\T{q}{}/2};\mu_Q;Q^2) \no 
&{} \qquad = f(\xbj,\T{k}{}-\T{q}{}/2;\mu_Q;Q^2) D^\text{pert}(\zh,\zh \T{q}{};\mu_Q;Q^2) + \order{\frac{m^2}{\Tscsq{q}{}}} \, .
\label{e.dpert_approx}
\end{align}
If we specialize to the order-$\alpha_s$ case, then the collinear perturbative expression from \eref{candidateqff_pert}, appropriate to the $\Tsc{k}{} \approx Q$ region, may be used for $D^\text{pert}(\zh,\zh \T{q}{};\mu_Q;Q^2)$. At order-$\alpha_s^n$, higher order versions may be used. Likewise, if we consider the region where 
\begin{equation}
\T{k}{} = -\T{q}{}/2 + \order{m} \, , 
\end{equation}
then 
\begin{align}
&{} f(\xbj,\T{k}{}-\T{q}{}/2;\mu_Q;Q^2) D(\zh,\zh \parz{\T{k}{}+\T{q}{}/2};\mu_Q;Q^2) \no 
&{} \qquad = f^\text{pert}(\xbj,-\T{q}{};\mu_Q;Q^2) D(\zh,\zh \parz{\T{k}{}+\T{q}{}/2};\mu_Q;Q^2) + \order{\frac{m^2}{\Tscsq{q}{}}} \, ,
\label{e.fpert_approx}
\end{align}
where $f^\text{pert}(\xbj,-\T{q}{};\mu_Q;Q^2)$ is the $n^\text{th}$-order perturbative expression appropriate to $\Tsc{k}{} \approx Q$. 
Again, when we specialize to the order-$\alpha_s$ treatment, the perturbative expression in \eref{candidateqpdf_pert} can be used, and at order-$\alpha_s^n$, the higher order versions of these expressions may be used. 

Having the expansions in \eref{dpert_approx} and \eref{fpert_approx} on hand motivates us to rewrite \eref{rewrite} in the form 
\begin{align}
\label{e.add_subtract}
\left[ f, D \right] 
&{}= D(\zh,\zh \T{q}{};\mu_Q;Q^2) \parz{2 \pi \int_0^{\mu_Q} \diff{\Tsc{k}{}}{} \Tsc{k}{} f(\xbj,\T{k}{};\mu_Q;Q^2)} \no \qquad 
&+ f(\xbj,-\T{q}{};\mu_Q;Q^2) \parz{2 \pi \int_0^{\mu_Q} \diff{\Tsc{k}{}}{} \Tsc{k}{} D(\zh,\zh \T{k}{};\mu_Q;Q^2)} \no 
&{}+  \int \diff{^2 \T{k}{}}{} \left\{  f(\xbj,\T{k}{}-\T{q}{}/2;\mu_Q;Q^2) D(\zh,\zh \parz{\T{k}{}+\T{q}{}/2};\mu_Q;Q^2) 
 \right. \no
&{} \left. \qquad \qquad - D(\zh,\zh \T{q}{};\mu_Q;Q^2) f(\xbj,\T{k}{}-\T{q}{}/2;\mu_Q;Q^2) \Theta(\mu_Q - |\T{k}{} - \T{q}{}/2|) \right. \no &{} \left. \qquad \qquad -  D(\zh,\zh (\T{k}{}+\T{q}{}/2);\mu_Q;Q^2) f(\xbj,-\T{q}{};\mu_Q;Q^2) \Theta(\mu_Q - |\T{k}{} + \T{q}{}/2|) \right\} \, ,
\end{align}
where we have simply added and subtracted the first two lines from the exact \eref{rewrite} to get the integral on the last three lines. 
On the first two lines of \eref{add_subtract}, we may replace $f(\xbj,\T{q}{};\mu_Q;Q^2)$ and $D(\zh,\zh \T{q}{};\mu_Q;Q^2)$ by their perturbative collinear approximations from \eref{dpert_approx} and \eref{fpert_approx}. Since they are evaluated at $\T{q}{} \approx Q$, this only introduces power-suppressed errors. We may also identify the cutoff integrals on the first two lines with the cutoff definitions of the collinear pdfs and ffs in \erefs{fc_deff}{dc_deff}.
The integrand of the last three lines is suppressed by $\order{m^2/\Tscsq{q}{}}$ in regions where $\T{k}{} = \pm \T{q}{}/2 + \order{m}$. Therefore, we may restrict our consideration of its behavior to regions where 
\begin{align}
|\T{k}{1}| &=  |\T{k}{}-\T{q}{}/2| \sim \Tsc{q}{} \, ,\\
|\T{k}{2}| &=  |\T{k}{}+\T{q}{}/2| \sim \Tsc{q}{} \, , 
\end{align}
i.e., where both $\Tsc{k}{1}$ and $\Tsc{k}{2}$ are an order unity fraction of $\Tsc{q}{}$. Then, all TMD pdfs and ffs in the integrand of the last three lines of \eref{add_subtract} can be expanded in powers of $m^2/\Tscsq{q}{}$ and replaced by their perturbative approximations, with only power suppressed corrections. We thus have 
\begin{align}
\label{e.add_subtract2}
\left[ f, D \right]
&{}= D^\text{pert}(\zh,\zh \T{q}{};\mu_Q;Q^2) f^c(\xbj;\mu_Q) + \frac{1}{\zh^2} f^\text{pert}(\xbj,-\T{q}{};\mu_Q;Q^2) 
d^c(\zh;\mu_Q)\no 
&{}+  \int \diff{^2 \T{k}{}}{} \left\{  f^\text{pert}(\xbj,\T{k}{}-\T{q}{}/2;\mu_Q;Q^2) D^\text{pert}(\zh,\zh \parz{\T{k}{}+\T{q}{}/2};\mu_Q;Q^2) 
 \right. \no
&{} \left. \qquad \qquad - D^\text{pert}(\zh,\zh \T{q}{};\mu_Q;Q^2) f^\text{pert}(\xbj,\T{k}{}-\T{q}{}/2;\mu_Q;Q^2) \Theta(\mu_Q - |\T{k}{} - \T{q}{}/2|) \right. \no &{} \left. \qquad \qquad -  D^\text{pert}(\zh,\zh (\T{k}{}+\T{q}{}/2);\mu_Q;Q^2) f^\text{pert}(\xbj,-\T{q}{};\mu_Q;Q^2) \Theta(\mu_Q - |\T{k}{} + \T{q}{}/2|) \right\} + \order{\frac{m^2}{\Tscsq{q}{}}} \, \no
&{}= \left[ f, D \right]_{\text{ASY}} + \order{\frac{m^2}{\Tscsq{q}{}}}
\end{align}
Dropping the $\order{m^2/\Tscsq{q}{}}$ errors gives the 
asymptotic term that we sought. We will denote this ``asymptotic" approximation by $\left[ f, D \right]_{\text{ASY}}$, as indicated on the last line. 
It is calculable entirely within collinear perturbation theory, and it is an increasingly accurate approximate of the full cross section as $\Tsc{q}{} \propto Q$ and $Q \to \infty$. The derivation above of \eref{add_subtract2} applies at any order of $\alpha_s$, although for this paper we will be mostly interested in $\order{\alpha_s}$ expressions. 

Notice that it is the \emph{cutoff} definitions, 
\erefs{fc_deff}{dc_deff}, for the collinear functions, and not the usual $\msbar$ definitions, that appear on the first line of \eref{add_subtract2}. One recovers the full asymptotic term for the cross section by substituting this into \eref{hadrotens}.

To specialize to the $\order{\alpha_s}$ case at an input scale $Q=Q_0$, with the parametrizations in \erefs{candidateqff}{candidateqpdf}, one substitutes the expressions from \erefs{candidateqff_pert}{candidateqpdf_pert}. Equations~\eqref{e.cutpdf_param} and \eqref{e.cutff_param} are to be used for the $f^c(\xbj;\mu_{Q_0})$ and $d^c(\zh;\mu_Q)$ on the first line of \eref{add_subtract2}. If we drop $\order{\alpha_s^2}$ and $\order{m^2/Q^2}$ errors, the first line then exactly matches the more standard form of the $\order{\alpha_s}$ asymptotic term (see, e.g., Ref.~\cite{Nadolsky:1999kb}). 

The integral that starts on the second line of \eref{add_subtract2} is only non-zero at $\order{\alpha_s^2}$ or higher, so it may be dropped in a strictly $\order{\alpha_s}$ treatment. 
However, there are several advantages to retaining it. 
One is simply that it guarantees that, for $Q = Q_0$, we recover the exact asymptotic $\Tsc{k}{} \to Q_0$, $m/Q_0 \to 0$ limit of the order-$\alpha_s^n$ TMD-term. Another is that it ensures cutoff-invariance through the lowest non-trivial order. Recall that the cutoff-defined pdfs and ffs can in general use a cutoff $\mu_f$ that differs from $\mu_Q$.
In \eref{add_subtract2}, $\mu_f$ dependence would appear in $f^c$, $d^c$, and the $\Theta$ functions in the integral of the last three lines.
Dependence on $\mu_f$ enters the standard asymptotic term at order $\alpha_s^2$, but keeping the third term in \eref{add_subtract2} ensures that $\mu_f$ dependence enters $\left[ f, D \right]_{\text{ASY}}$ only at order $\alpha_s^3$.

\section{Example input scale treatment}
\label{s.matching}

Now we turn to demonstrating how the HSO treatment described in \srefs{sidis}{models} works in practice with explicit numerical implementations. 
Our purpose here is to compare the HSO treatment described thus far with the conventional steps for constructing phenomenological parametrizations, and to illustrate the improvements that are gained from using the former.

In \sref{basic_setup} below, we will summarize the basic formulas and in \sref{largeandsmallqT} we will review the usual decomposition of a transverse momentum dependent cross section into a TMD term, an asymptotic term, and a $Y$-term. In \sref{conventional}, we will review the conventional style of implementing TMD factorization and show examples of the complications that can arise, some of which were already mentioned in the introduction, and in \sref{hso} we show how these are solved within the HSO approach.

 In our calculations, we focus on the TMD pdfs and ffs parametrized at an initial scale $Q=Q_0$, a scenario previously addressed in \cite{Boglione:2014oea}. 
 Estimating the lowest $Q_0$ for which TMD factorization remains valid is rather non-trivial~\cite{Gonzalez-Hernandez:2022ifv}, and we leave it as an open question. For purposes of illustration, we will try two values in sections \ref{s.conventional} and \ref{s.hso} below, from the relatively low (and reasonable) $Q_0 = 4.0$~GeV, to the (far too conservative) $Q_0 = 20.0$~GeV, to demonstrate how the procedure works for both a small and a large choices of $Q_0$. 

\subsection{Basic setup}
\label{s.basic_setup}
The standard expression for the SIDIS differential cross section in terms of the structure functions $F_1$ and $F_2$ is
\begin{align}
\label{e.csecexplicit}
\frac{\diff{\sigma}}{\diff{\xbj}\,\diff{y}\,\diff{\zh}\,\diff{\Tscsq{q}{}}}
={}&
\frac{\pi ^2 \alpha_\text{em}^2 \zh}{Q^2\, \xbj \, y}
\left[
F_1\,\xbj \,y^2
+
F_2\,(1-y)
\right]
 \, ,
\end{align}
where the $F$ structure functions are the usual ones obtained by contracting the projectors in \eref{F12proj} with the hadronic tensor. In the small-$\Tsc{q}{}$ approximation, the structure functions are expressed in terms of TMD pdfs and ffs,
\begin{align}
&F
={}
F^{\text{TMD}}+\order{m/Q,\Tsc{q}{}/Q} \, ,
\label{e.F12smallqT}
\\
\no
F_{1}^{\text{TMD}}
\equiv{}
2\,\zh\,\sum_{j}& |H|^2_j \left[ f_{j/p}, D_{h/j} \right]\, ,
\qquad
F_{2}^{\text{TMD}}
\equiv
4\,\zh \,\xbj\,\sum_{j} |H|^2_j \left[ f_{j/p}, D_{h/j} \right]\,,
\label{e.F12tmd}
\end{align}
where the ``$\text{TMD}$'' superscript denotes the small-$\Tsc{q}{}$ approximation. Compare \eref{F12tmd} with \eref{hadrotens} for the hadronic tensor.
We will use the $O(\alpha_s)$ hard factor $|H|^2_j$ from \eref{hardpart} in any calculations below.
Calculating \eref{F12tmd} in a specific phenomenological implementation involves making choices about how to parametrize the TMD functions $f_{i/p}$ and $D_{h/j}$, including choices about nonperturbative models and/or calculations at the input scale, the order of precision in perturbative parts, and any other approximations or assumptions used in the construction of a specific set of parametrizations. 

\subsection{Combining large ($F^\text{FO}$) and small ($F^\text{TMD}$) transverse momentum calculations}
\label{s.largeandsmallqT}

Before we contrast the $F^{\text{TMD}}$ calculations in the conventional and HSO styles, let us review the usual steps  
for merging calculations done with TMDs with purely collinear factorization calculations designed for the $\Tsc{q}{} \approx Q$ region.

In the region where $\Tsc{q}{}\approx Q$, the approximations in \eref{F12tmd} fail. However, this is the region where fixed-order collinear factorization calculations, which use ordinary collinear pdfs and ffs, are most reliable. 
We express the large-$\Tsc{q}{}$ fixed-order collinear approximation to the structure functions as
\begin{align}
F
={}&
F^{\text{FO}}+\order{m/\Tsc{q}{}}\,,
\qquad
F^{\text{FO}}={}
 \sum_{i,j} d_{B/i} \otimes \hat{F}_{ij}\otimes f_{j/p}  \,,
\label{e.F12fo}
\end{align}
where the indices $i,j$ run over parton flavors, and the $\text{FO}$ superscript stands for ``fixed-order.'' 
A choice must be made for the UV scheme that defines the collinear functions $f_{i/p}$ and $D_{h/j}$.
The most common is renormalization in the $\msbar$ scheme. 
The $\hat{F}_{ij}$  are the partonic versions of the structure functions, and they have been calculated up to at least $O(\alpha_s^2)$  \cite{Aurenche:2003by,Daleo:2004pn,Wang:2019bvb}. In our calculations, we will use $O(\alpha_s)$ results \cite{Nadolsky:1999kb,Koike:2006fn,Anselmino:2006rv}.

Following standard conventions, we will use the phrase ``fixed order cross section'' as a short hand for \eref{csecexplicit} calculated with the large-$\Tsc{q}{}$ approximation in \eref{F12fo}.\footnote{Note that the asymptotic term of \sref{largeqtasymptote} is also calculated in fixed order perturbation theory. However, in the terminology of this section ``fixed order term'' applies specifically to calculations done using the non-asymptotic \eref{F12fo}.}  
While $F^{\text{TMD}}$ gives an accurate treatment of the $\Tsc{q}{} \approx m$ region, and $F^{\text{FO}}$ provides an accurate treatment of the $\Tsc{q}{} \approx Q$ region, what is ultimately needed is a factorized expression with only $\order{m^2/Q^2}$-suppressed errors point-by-point in $\Tsc{q}{}$. To construct it systematically, one starts by writing the structure functions in the TMD (low-$\Tsc{q}{}$) approximation with the error term made explicit, 
\begin{equation}
F = F^{\text{TMD}} + \left[F - F^{\text{TMD}} \right] \, . \label{e.errerterm1}
\end{equation}
The error term in braces is only unsuppressed when $\Tsc{q}{}$ is large relative to $m$. Thus, it can be calculated in collinear factorization with only $m^2/\Tscsq{q}{}$-suppressed errors. Since the error term itself is $\order{\Tscsq{q}{}/Q^2}$, the result is that the overall error is $m^2/Q^2$-suppressed point-by-point in $\Tsc{q}{}$. Thus, we define
\begin{equation}
\lim_{m/\Tsc{q}{} \to 0} F^{\text{TMD}} = F^{\text{ASY}}
\end{equation}
to be the $\Tsc{q}{} \sim Q$, $Q \to \infty$ asymptote of the TMD approximation, as it is calculated in fixed order collinear factorization. The ``$\sim$'' means the ratio $\Tscsq{q}{}/Q^2$ is to be held fixed as $Q \to \infty$.  Applied to \eref{errerterm1}, the structure function becomes 
\begin{equation}
F = F^{\text{TMD}} + \left[ F^{\text{FO}} - F^{\text{ASY}} \right] \, + \order{m^2/Q^2}  \label{e.errerterm2} \, .
\end{equation}
The asymptotic term is consctructed to accurately describe the $m \ll \Tsc{q}{} \ll Q$ region -- both $\Tsc{q}{} \ll Q$ and $m \ll \Tsc{q}{}$ approximations have been applied simultaneously. For this paper, this is simply \eref{add_subtract2} applied to structure functions. 

A minor subtlety is that the exact form of the asymptotic term $F^{\text{ASY}}$ depends on the details of how collinear pdfs and ffs are defined and on how higher order corrections in the perturbative expansion are truncated. If, in an $\order{\alpha_s^n}$ calculation, for example, the cutoff-defined pdfs and ffs of \eref{add_subtract2} are replaced by their corresponding $\msbar$ definitions, then the resulting asymptotic terms will generally differ by $\order{m^2/Q^2}$-suppressed and $\order{\alpha_s^{n+1}}$-suppressed amounts.  Furthermore, while $F^{\text{ASY}}$ is in principle equal to the low-$\Tsc{q}{}$ limit of $F^{\text{FO}}$ as $Q \to \infty$, generally this is only exactly true in calculations at the working order of perturbation theory. In calculations at a fixed $Q$, the two asymptotic terms will typically differ by higher-order $\alpha_s$ and power-suppressed terms. In other words, if $F^{\text{ASY}}$ is calculated to $\order{\alpha_s^n}$ with the cutoff scheme for pdfs and ffs, and $F^{\text{FO},r}$ is calculated to the same order in some other scheme $r$, then one will generally find
\begin{equation}
\left[ \lim_{\Tsc{q}{}/Q \to 0} F^{\text{FO},r} \right]^{\order{\alpha_s^n}} - 
\left[ F^{\text{ASY
}} \right]^{\order{\alpha_s^n}} = \order{\alpha_s^{n+1},m^2/Q^2} \, . \label{e.asypmptoticcompare}
\end{equation}
That is, there is a family of valid schemes for defining the exact asymptotic term at a given order, though some schemes can be preferable to others in the context of minimizing errors.  Indeed, it is the first term in \eref{asypmptoticcompare}, with $r=\msbar$, that represents the most common approach used in the past for calculating the asymptotic term.
We will call the asymptotic term calculated using \eref{add_subtract2} $F_{\text{HSO}}^{\text{ASY}}$.

Together, the second two terms in \eref{errerterm2} are often called the ``$Y$-term,'' and the structure function is written as
\begin{equation}
F = F^{\text{TMD}} + Y \, + \order{m/Q}  \label{e.errerterm3} \, .
\end{equation}
to emphasize the role of $Y$ as a large-$\Tsc{q}{}$ correction to calculations done with TMD pdfs and ffs. Of course, the precise value of the $Y$-term contribution depends on the specific version of the asymptotic term.

In conventional treatments, the fixed order term is calculated with collinear functions in the $\msbar$ scheme. The specific version of the asymptotic structure functions used is the first term in \eref{asypmptoticcompare}, so that 
\begin{align}
\label{e.stdef}
F_{\text{ST}}^{\text{FO}}
=&
F^{\text{FO},\msbar},\qquad
F_{\text{ST}}^{\text{ASY}} 
=
\lim_{\Tsc{q}{}/Q \to 0} F^{\text{FO},\msbar} \, ,
\end{align}
with ``$\text{ST}$'' subscripts to indicate ``standard.''
We will call a calculation of the asymptotic term done in the style of \sref{largeqtasymptote} $F_{\text{HSO}}^{\text{ASY}}$ to distinguish it from \eref{stdef}. Since $F_{\text{HSO}}^{\text{ASY}}$ is calculated with cutoff definitions for the collinear pdfs and ffs, this suggests that the cutoff definitions might be preferred as well for calculating $F^{\text{FO}}$. However, switching between the $\msbar$ and cutoff schemes in $F^{\text{FO}}$ only produces power suppressed and perturbative errors beyond the working order in $\alpha_s$. Therefore, one may consistently interchange cutoff and $\msbar$ definitions, and we will use $F_{\text{ST}}^{\text{FO}}$ for our calculation of the fixed order structure function. We will see in later sections that the effect of switching between the two is small relative to the overall improvements from using the HSO approach. An interesting question for the future is whether calculations of $F^{\text{FO}}$ can be improved by switching to a cutoff scheme for the collinear functions, but we leave this to future work.

\subsection{The TMD term in the conventional treatment}
\label{s.conventional}

The usual approach to applying TMD factorization to phenomenology has been reviewed in many places, so we will not repeat the details here. Readers are referred to, for example, Refs.~\cite{Collins:2011qcdbook,Collins:2014jpa,Rogers:2015sqa} and references therein. The standard expression used in calculations follow from making the following replacement in Eqs.~(\ref{e.F12tmd}):
\begin{align}
\left[ f_{j/p}, D_{h/j} \right] &{}\to 
    \int \frac{\diff[2]{\T{b}{}}}{(2 \pi)^2}
    ~ e^{-i\T{q}{}\cdot \T{b}{} }
    ~ \tilde{f}_{j/p}^{\text{OPE}}(\xbj,\bstar;\mubstar,\mubstar^2) 
    ~ \tilde{D}_{h/j}^{\text{OPE}}(\zh,\bstar;\mubstar,\mubstar^2)\no
    &{} \times \exp
    \left\{ 2 \int_{\mubstar}^{\mu_{Q}} \frac{d \mu^\prime}{\mu^\prime} \left[\gamma(\alpha_s(\mu^\prime);1) 
- \ln \frac{Q}{\mu^\prime} \gamma_K(\alpha_s(\mu^\prime))
  \right] +\ln \frac{Q^2}{\mubstar^2} \tilde{K}(\bstarsc;\mubstar)
  \right\}
  \no
&{} \times \exp\left\{
-g_{j/p}(\xbj,\Tsc{b}{}) - g_{h/j}(\zh,\Tsc{b}{}) - g_K(\Tsc{b}{}) \ln \parz{\frac{Q^2}{Q_0^2}} \right\} \, . \label{e.replaceWtmd1}
\end{align}
The $\tilde{f}_{j/p}^{\text{OPE}}$ and $\tilde{D}_{h/j}^{\text{OPE}}$ on the first line are the TMD pdfs and ffs in $\Tsc{b}{}$-space, expanded and truncated in an operator product expansion. The $\gamma$, $\gamma_K$ and $\tilde{K}$ are the usual evolution kernels. The ``$\bstar$'' method has been used to regulate  $\tilde{f}_{j/p}^{\text{OPE}}$, $\tilde{D}_{h/j}^{\text{OPE}}$, and $\tilde{K}$ at large $\Tsc{b}{}$. (See reviews of the $\bstar$ method in Sec.~IXA of \cite{Gonzalez-Hernandez:2022ifv} and in Sec.~VIII of \cite{Aslan:2022zkz}.) The most common choice for a functional form for $\bstar$ is 
\begin{align}
\label{e.bstar}
  \bstar(\Tsc{b}{}) = \frac{ \T{b}{} }{ \sqrt{ 1 + \Tscsq{b}{}/\bmax^2} } \, ,
\end{align}
where $\bmax$ is a transverse size scale that demarcates a separation between large and small transverse size regions. In principle, both the functional form of \eref{bstar} and the value of $\bmax$ are completely arbitrary, but a small $\bmax$ justifies the use of the OPE on the first line of \eref{replaceWtmd1}; the error term in the approximation in \eref{replaceWtmd1} is suppressed by powers of $m \,\bmax$.
All of the nonperturbative transverse momentum dependence is contained in the $\Tsc{b}{}$-space functions $g_{j/p}$, $g_{h/j}$, and $g_K$, whose definitions in terms of the more fundamental correlation functions are 
\begin{align}
-g_{j/p}(x,\Tsc{b}{}) &{} \equiv \ln \parz{ \frac{\tilde{f}_{j/p}(x,\T{b}{};\mu_{Q_0},Q_0^2)}{\tilde{f}_{j/p}(x,\bstar;\mu_{Q_0},Q_0^2)}}\, ,
\qquad
-g_{h/j}(z,\Tsc{b}{}) {} \equiv \ln \parz{ \frac{\tilde{D}_{h/j}(z,\T{b}{};\mu_{Q_0},Q_0^2)}{\tilde{D}_{h/j}(z,\bstar;\mu_{Q_0},Q_0^2)}}\,,
\label{e.gdef}
\end{align}
and
\begin{equation}
\label{e.gKdef}
  g_K(\Tsc{b}{}) 
  \equiv \tilde{K}(\bstarsc;\mu) -\tilde{K}(\Tsc{b}{};\mu) \, .
\end{equation}

Conventional methods replace each of the $g$-functions, $g_{j/p}$, $g_{h/j}$, and $g_K$, by an ansatz, with parameters to be fitted from measurements. The simplest and most common choices (e.g.~\cite{Davies:1984sp,Balazs:1995nz,Landry:1999an}) are based on simple power laws like 
\begin{align}
g_{j/p}(\xbj,\Tsc{b}{})={}&
\frac{1}{4}\,M_{F}^2\,\Tscsq{b}{}\, ,
\qquad 
g_{h/j}(\zh,\Tsc{b}{}) ={}
\frac{1}{4\,\zh^2}M_{D}^2\,\Tscsq{b}{}\, 
\label{e.gmodel}
\end{align}
for the input nonperturbative functions, where $M_{F}$ and  $M_{D}$ are fit parameters. 
For the CS kernel, common parametrizations are
\begin{equation}
g_K(\Tsc{b}{}) = \frac{1}{2} M_{K}^2\Tscsq{b}{}\,  \qquad \text{or}  \qquad 
g_K(\Tsc{b}{}) = \frac{g_2}{2\,M_{K}^2} \ln\parz{1 + M_{K}^2 \Tscsq{b}{}} \, , \label{e.gkmodels}
\end{equation}
where $M_{K}$ and $g_2$ are fit parameters. 
The first of these functional forms is common in typical applications, but it conflicts with the expectation that evolution is slow at moderate $Q$ \cite{Sun:2013dya,Sun:2013hua}. As a result, it was suggested in Ref.~\cite{Collins:2014jpa} that $g_K(\Tsc{b}{})$ should exhibit very nearly constant behavior at large $\Tsc{b}{}$, a behavior closely modeled by a logarithmic function. More complex fit parametrization ansatzes for all the g-functions have been introduced more recently (see for instance Refs.~\cite{Bury:2022czx,Bacchetta:2022awv}), but the general approach of taking combinations of simple functional forms that reduce to power law behavior at small $\Tsc{b}{}$ is similar to the above.

Note that, in the $\bstarsc$-approach, before any truncation approximations are made, the product of TMD correlation functions must satisfy 
\begin{equation}
\frac{\diff{}{}}{\diff{\bmax}{}} \left[ f_{j/p}, D_{h/j} \right] = \order{m \bmax} \, . \label{e.bmaxind}
\end{equation}
That is, dependence on $\bmax$ or on the form of $\bstar(\Tsc{b}{})$ must be a negligible power correction for reasonably small $\bmax$.\footnote{The power-suppressed errors on the right side of \eref{bmaxind} will typically be $m^2 \bmax^2$, but the precise power of the suppression is not important for our present discussion.} In calculations at a specific order in $\alpha_s$, violations of \eref{bmaxind} may enter only through neglected higher orders in $\alpha_s$. A significant violation of \eref{bmaxind} in a TMD parametrization may indicate either that higher orders need to be included, or that $\bmax$ has been chosen to be too large. 
A failure to find a negligible right side of \eref{bmaxind} is thus a useful diagnostic tool. 

We will label structure functions calculated in the conventional approach by $F^{\text{TMD}}_{\text{ST}}$, with ``$\text{ST}$'' for ``standard,'' and we will use this notation 
regardless of whichever specific model is used for the $g$-functions. What makes an approach ``conventional'' in the sense that we mean in this paper is that it imposes no extra, additional constraints on the $g$-functions to ensure consistent matching with collinear factorization.  
Specifically, the ansatzes of traditional approaches do not explicitly enforce the integral connection between collinear and TMD pdfs and ffs in \eref{cutoff_def}, or guarantee a smooth interpolation to the large $\Tsc{k}{}$ collinear factorization region. 

In the following numerical examples, 
we will use {\tt CTEQ6.6}  pdfs~\cite{Nadolsky:2008zw} (central values) and {\tt MAPFF1.0} ffs for $\pi^+$~\cite{Khalek:2021gxf} (avergage over replicas), implemented in {\tt LHAPDF6}~\cite{Buckley:2014ana}. 
We postpone a more detailed analysis that includes the uncertainty associated with the chosen {\tt LHAPDF6} sets for a later publication. For the purpose of this paper, we effectively assume ``complete knowledge" of the collinear pdfs and ffs in the $\msbar$ scheme stressing that our main points, and the logic behind the HSO approach, are not affected by such choices.
The left-hand panels of \fref{tmdg_lines} show the differential SIDIS cross section for $Q_0 = 4.0$~GeV within the various different approximations discussed in \sref{basic_setup} and \sref{largeandsmallqT}, including the $F^{\text{TMD}}_{\text{ST}}$ (the TMD approximation), the $F^{\text{FO}}_{\text{ST}}$ ($\Tsc{q}{} \approx Q$ approximation), and the $F^{\text{ASY}}_{\text{ST}}$ (asymptotic term) calculations. 
We use $x=0.1$, $z=0.3$ and $y=0.5$, which are kinematics accessible to both the COMPASS experiment \cite{COMPASS:2017mvk} and the EIC \cite{AbdulKhalek:2021gbh}. To emphasize alternately the large-$\Tsc{q}{}$ and small-$\Tsc{q}{}$ regions, we have plotted the curves on a logarithmic scale in the upper left panel and a linear scale in the lower left panel.
We take the $g$-functions to be parametrized as in \eref{gmodel}, and the RG scale is $\mu_{Q_0}=Q_0$. 
The curves are the TMD (solid red line), fixed order (dot-dashed black line) and asymptotic (dashed blue line) terms.
Despite the small values used for the mass parameters, $M_{\text{F}}=M_{\text{D}}/z=0.1\,\text{GeV}$, the asymptotic term is nowhere close to overlapping with either the TMD or the fixed order terms anywhere in the range of $\Tsc{q}{}$ between $0$ and $4\,\text{GeV}$. This is a violation of the consistency requirement that, with a  sufficiently large input scale $Q_0$, there must be a region $\Lambda_\text{QCD} \ll \Tsc{q}{}\ll Q_0$ where the asymptotic term is simultaneously a good approximation of both the TMD and the $\Tsc{q}{} \approx Q_0$ fixed order cross sections. This is a complication that arises frequently in the conventional methodology, and it is one that we alluded to in \sref{intro}. Among the reasons for the mismatch is a failure to impose the integral relation in \eref{cutoff_def} directly upon the $g$-functions in \eref{gmodel}.

One might suspect that the mismatch is a consequence of the input scale $Q_0$ being too small. To test this, we also consider the same computation, using the same nonperturbative mass scales, but now with an unreasonably large input scale of $Q_0=20\,\text{GeV}$. The result is shown in the right-hand panels of~\fref{tmdg_lines}. Again, the upper panel is on a logarithmic scale, while the lower panel uses the linear scale to emphasize the region of smaller $\Tsc{q}{}$. 
The agreement between the asymptotic and TMD terms improves, but even here there is a startlingly large mismatch between the three calculations in the region where $\Tsc{q}{}$ is small but comparable to $Q_0$. Even for $Q_0 \approx 20$~GeV, there is no region of $\Tsc{q}{}$ where the three curves overlap simultaneously to a satisfactory degree. 
This point is made especially clear in the linear scale plots. 

Note that this complication is independent of evolution or the question about how many orders of logarithms of $Q/\Tsc{q}{}$ should be resummed. If the connection to collinear factorization is to be consistent, there must be a region where $\Tsc{q}{}$ is a fixed fraction of $Q$ and all three calculations merge in the limit as $Q \to \infty$.  Moreover, for any $Q$ where we expect TMD factorization to be valid, the TMD and asymptotic terms should at least approximately match one another when $\Tsc{q}{}$ is comparable to $Q$. It is a contradiction, then, if this fails at the input scale. 
Note that the mismatches, both quantitative and qualitative, between the TMD terms and their expected asymptotic behavior is especially visible in the lower panels where the curves are plotted with linear axes.

For generating the plots in \fref{tmdg_lines}, it was necessary to fix the mass scales $M_F$ and $M_D$ in \eref{gmodel}. The observed trends are quite general, however, and to demonstrate this we show the same $Q_0 = 4.0$~GeV calculation in the left-hand panel of \fref{tmdg_bands}, but now with bands representing ranges of typically-sized nonperturbative mass scales,
\begin{align}
&0.1~\text{GeV} \leq M_F \leq 0.4~\text{GeV} \, , \label{e.range1} \\
&0.1~\text{GeV} \leq M_D/z \leq 0.3~\text{GeV} \, . \label{e.range2}
\end{align}
The value of $\bmax$ for this plot remains fixed at $1.0$~GeV$^{-1}$. Even with the freedom to adjust these nonperturbative parameters, it is clear that it is not possible to achieve reasonable agreement between the TMD term and the asymptotic term, even 
in regions where $\Tsc{q}{}$ is comparable to $Q_0$. The TMD bands do touch the asymptotic curve at around $\Tsc{q}{} \approx 0.5$~GeV, but the two curves have very different qualitative shapes for all $M_D$ and $M_F$. For larger $\Tsc{q}{}$, there is no approximate agreement between the asymptotic and TMD terms, regardless of $M_F$ and $M_D$. Indeed, the TMD band departs from  the asymptotic term at around $\Tsc{q}{}\approx1.2\,\text{GeV}$.

Another way to see the problems with the conventional treatment here is to observe that the approximate $\bmax$-independence of  \eref{bmaxind} is very badly violated with typical values of $\bmax$, as shown by the right-hand panel in 
\fref{tmdg_bands}, which displays the TMD term with bands for $\bmax$ variations from the very small value of $0.1$~GeV$^{-1}$ up to a maximum typical value of $\bmax = 1.5$~GeV$^{-1}$ used in phenomenological applications.
The bands are with fixed mass scales of $M_F = M_D/z = 0.25$~GeV. The orders-of magnitude variation badly contradicts the original $\bmax$-independence that exists before the OPE approximations. It implies that the $M_F$ and $M_D$ parameters must be given their own $\bmax$-dependence
to (at least approximately) cancel the explicit $\bmax$-dependence seen in the figure. 
However, the far more modest $M_F$ and $M_D$ dependence seen in the left-hand panel shows that this cannot be made to work with typical model parametrizations of the $g$-functions and reasonable nonperturbative values for $M_F$ and $M_D$. 

As a consequence of the strong $\bmax$ sensitivity, practical phenomenological applications will often effectively promote $\bmax$ to the status of an extra nonperturbative parameter as opposed to treating it as an entirely arbitrary cutoff. That is, attempts to approximately preserve \eref{bmaxind} are effectively abandoned. But the result is that the large transverse momentum behavior becomes sensitive to parameters that are in principle to be restricted to describing only the nonperturbative small transverse momentum region. The predictive power that is gained from collinear factorization and the OPE is then compromised. This is a problem that has been well-known for some time~\cite{Qiu:2000hf}.

The above observations illustrate that nonpertubative transverse momentum dependence in the conventional methodology has an unacceptably large impact on the large transverse momentum region, in a way that violates consistency with collinear factorization.

\begin{figure}
     \centering
         \includegraphics[width=0.49\textwidth]{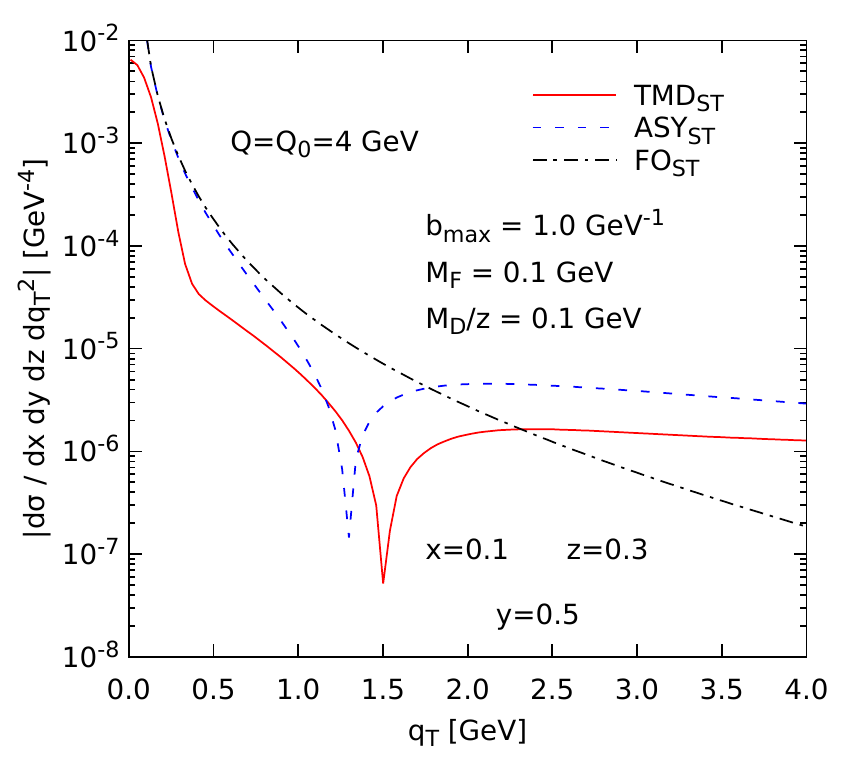}
         \includegraphics[width=0.49\textwidth]{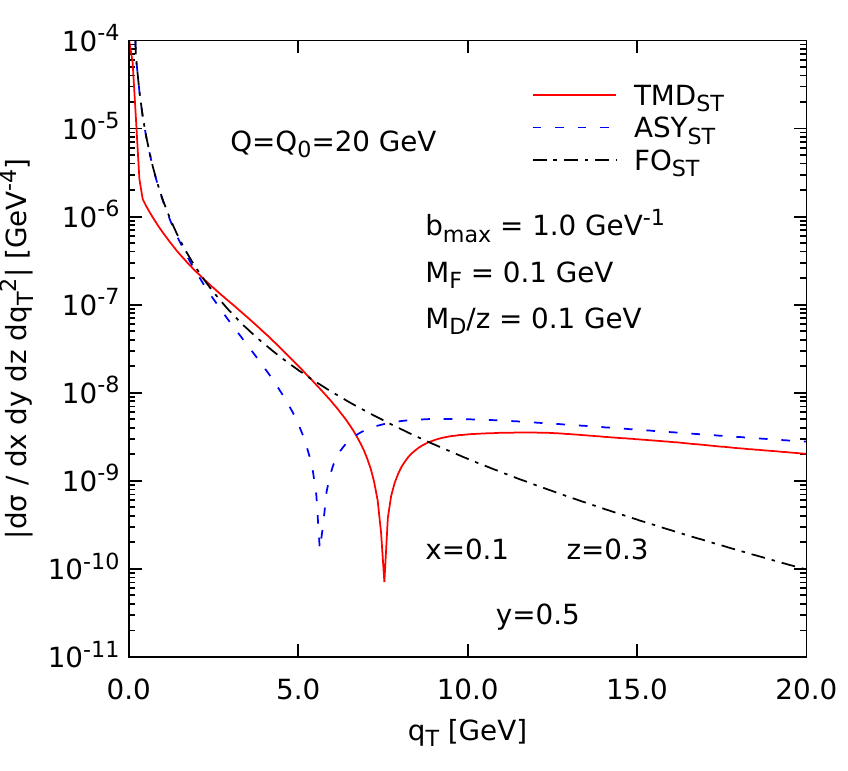}

         \includegraphics[width=0.49\textwidth]{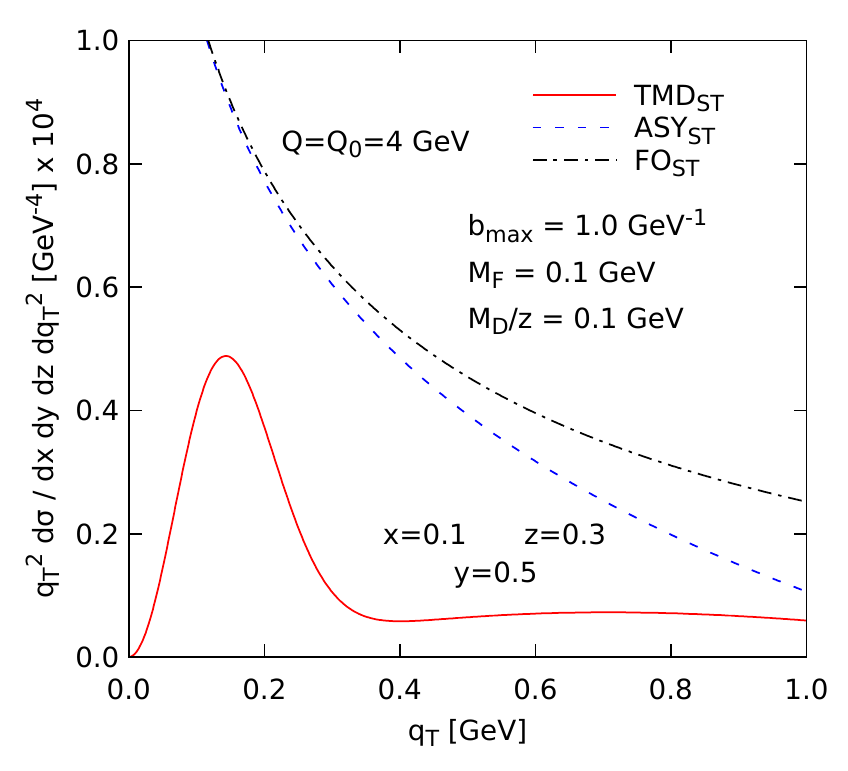}
         \includegraphics[width=0.49\textwidth]{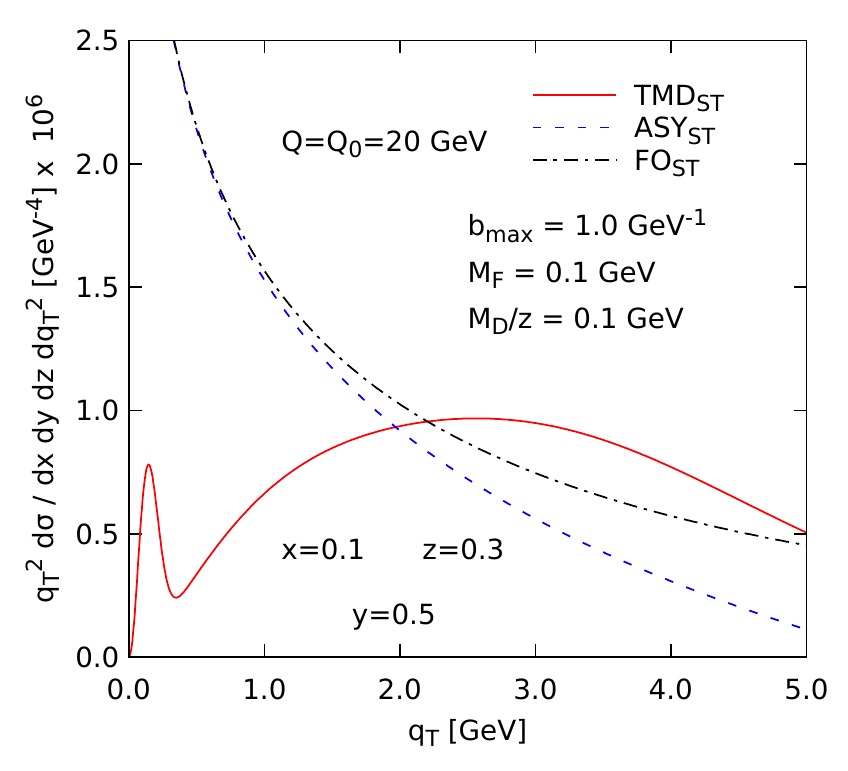}
        \caption{
        SIDIS differential cross section (absolute value) in the standard approach, within different approximations for the structure functions:  $F^{\text{TMD}}_{\text{ST}}$ (solid red line), $F^{\text{ASY}}_{\text{ST}}$ (dashed blue line) and  $F^{\text{FO}}_{\text{ST}}$ (dot-dashed black line). The chosen kinematics roughly correspond to regions accessible by the COMPASS experiment and the EIC. 
        The TMD term is calculated with the quadratic model for the g-functions of \eref{gmodel}, at fixed values for the small-mass parameters $M_{\text{F}}=M_{\text{D}}/z=0.1\,\text{GeV}$, and we have used the  $\bstarsc$ prescription of \eref{bstar} with $\bmax=1.0\,\text{GeV}^{-1}$. We consider the cross section at two values of the input scale $Q_0$, and no TMD evolution is performed. Left: The cross section is shown  for $Q_0=4.0\,\text{GeV}$. Right: The cross section is shown  for $Q_0=20.0\,\text{GeV}$. For visibility, the bottom panels show the same curves as the top, but with a vertical linear scale and a reduced range of $\Tsc{q}{}$. 
        Note that, despite the small values of the mass parameters, the three approximations never overlap in the intermediate region of transverse momentum, $m\ll \Tsc{q}{}\ll Q$.
        }
        \label{f.tmdg_lines}
\end{figure}
\begin{figure}
     \centering
         \includegraphics[width=0.49\textwidth]{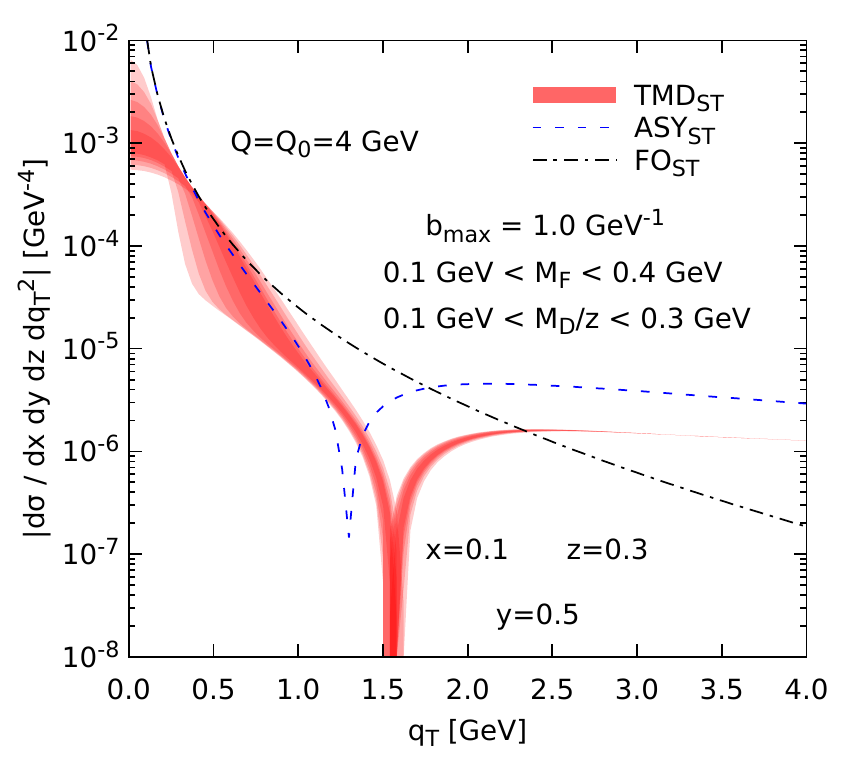}
         \includegraphics[width=0.49\textwidth]{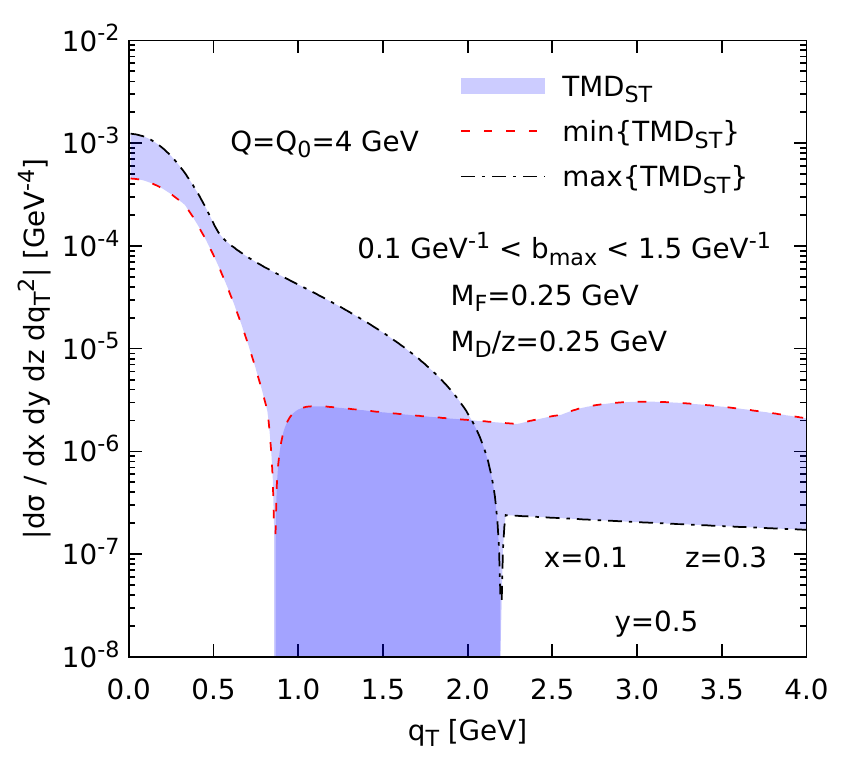}
        \caption{
        Variation of the TMD cross section (absolute value), in the standard approach, with respect to the small-mass parameters of \eref{gmodel} (left), and $\bmax$ (right). In both cases, we have chosen the same kinematics as in the left panel of \fref{tmdg_lines}, and we have set $Q=Q_0=4.0\,\text{GeV}$, and no TMD evolution is performed.  Left: the red band shows the envelope for the TMD term obtained by varying the model masses $M_{\text{F}}$ and $M_{\text{D}}$. Note the large variation of the band in the region where the asymptotic term (dashed blue line) and the fixed order term (dot-dashed black line) start to overlap, which results from the unconstrained behavior of the TMD term at large $\Tsc{q}{}$. At very large values of $\Tsc{q}{}$, the TMD and asymptotic terms are not consistent. Right:  Envelope showing the variation of the TMD term (blue band) with respect to the value of $\bmax$, at fixed values of the model masses. (Note that the edges of the envelope are not necessarily the curves associated with the extrema of the chosen range for $\bmax$). The strong $\bmax$ dependence results from the lack of constraints on the models for the g-functions in our example. This dependence persists  even in the region $\Tsc{q}{}\sim Q$, where the OPE  should in principle determine the behaviour of the TMD cross section. 
        }
        \label{f.tmdg_bands}
\end{figure}
\begin{figure}
     \centering
         \includegraphics[width=0.49\textwidth]{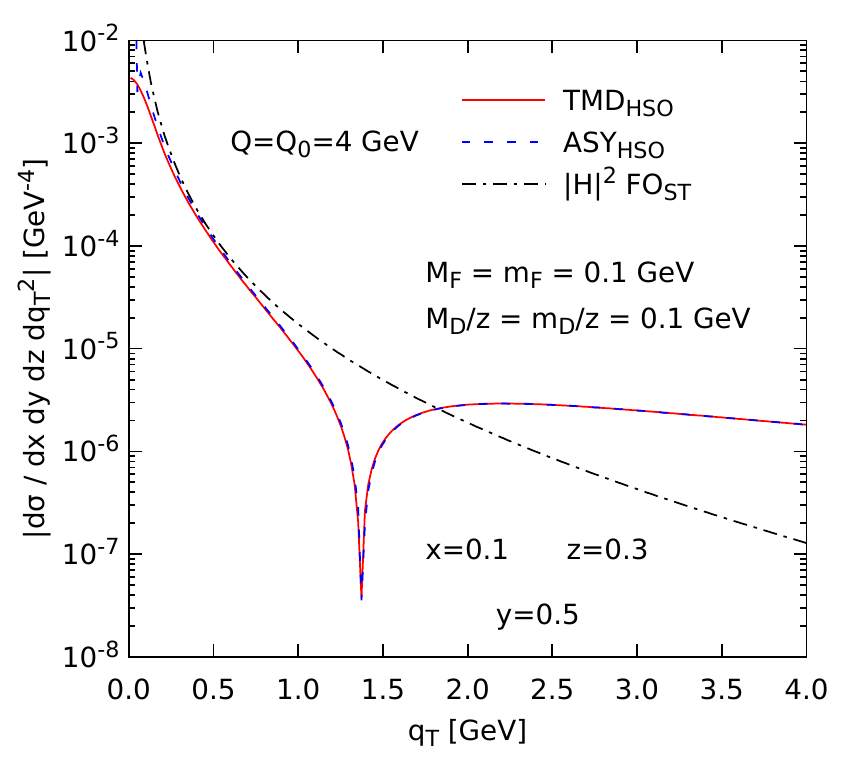}
          \includegraphics[width=0.49\textwidth]{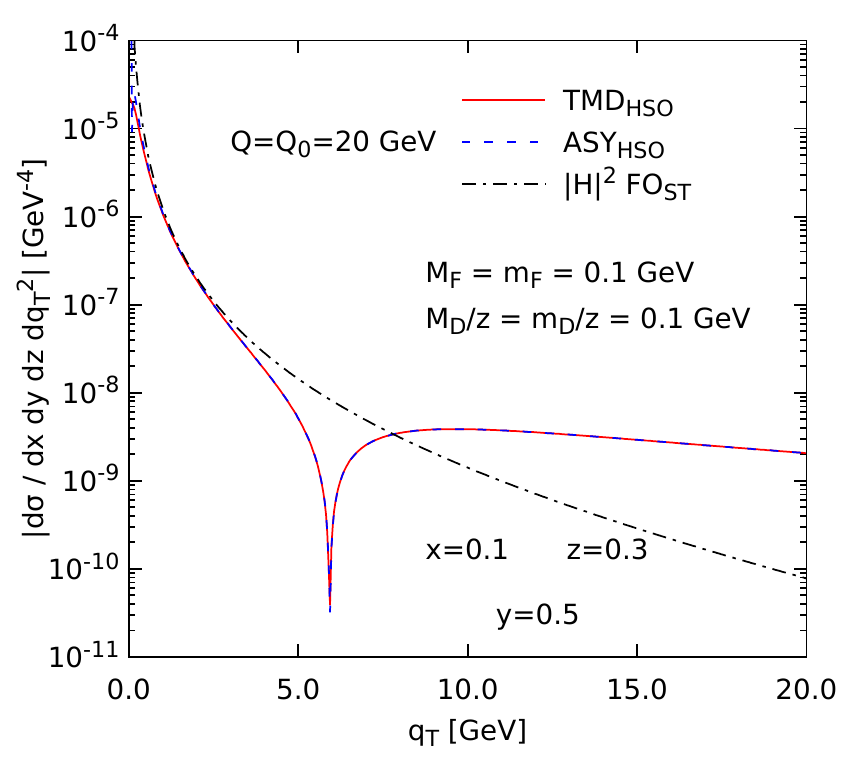}

          \includegraphics[width=0.49\textwidth]{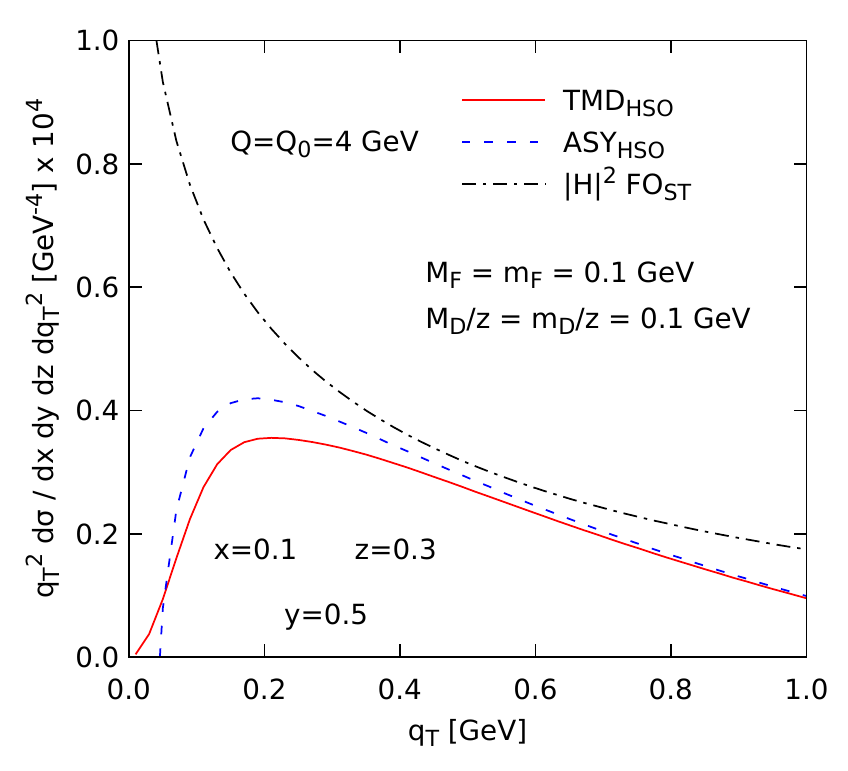}
          \includegraphics[width=0.49\textwidth]{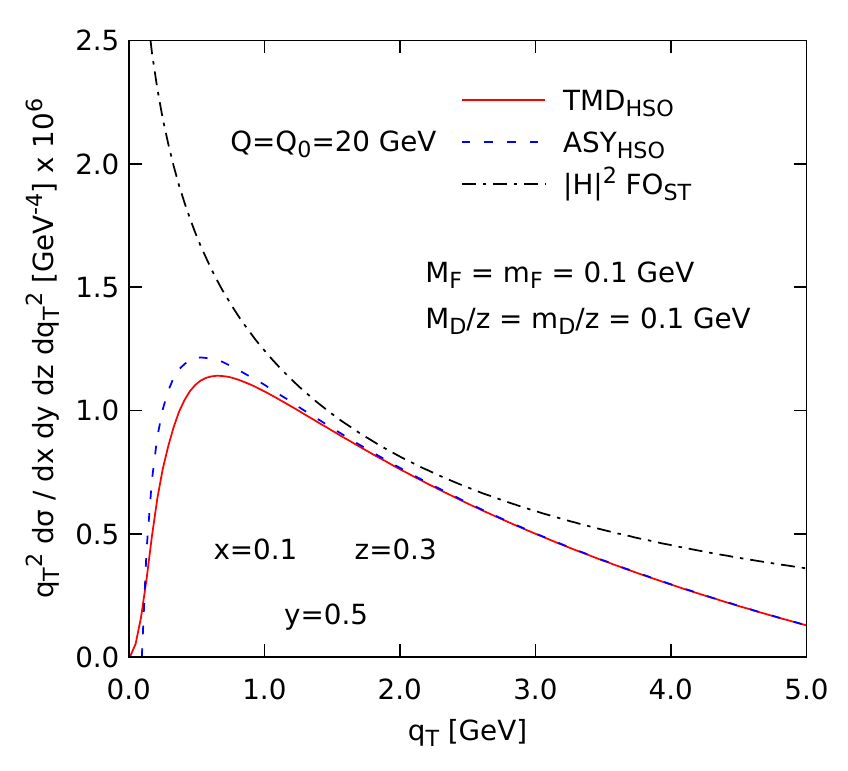}
          
        \caption{
        SIDIS differential cross section (absolute value) in the HSO approach, comparing different approximations for the structure functions:  $F^{\text{TMD}}_{\text{HSO}}$ (solid red line), $F^{\text{ASY}}_{\text{HSO}}$ (dashed blue line) and 
        $F^{\text{FO}}$ as in \eref{Ffohso}
        (dot-dashed black line). For comparison, the same kinematics have been used as in \fref{tmdg_lines}. 
        The TMD term is calculated with the Gaussian models of \erefs{npmodels}{npmodels_cutoff}, with appropriate constraints as in \eref{candidateqff} and \eref{candidateqpdf}. These models essentially determine the g-functions, similar to \eref{gmodel} in the standard approach, but with the correct treatment of the large-$\Tsc{k}{}$ behavior and the implementation of integral relations.
        To allow for a meaningful comparison, we use the same values for the small-mass parameters $M_{\text{F}}=M_{\text{D}}/z=0.1\,\text{GeV}$ as in \fref{tmdg_lines}. The masses appearing in \eref{candidateqff} are set to $m_{D_{h,j}}=m_{D_{h,g}}=M_{\text{D}}$, and those in \eref{candidateqpdf} to $m_{f_{i,p}}=m_{f_{g,p}}=M_{\text{F}}$.
        We compute the cross section at the same two values for the input scale $Q_0$ considered in \fref{tmdg_lines}. Left: The cross section for $Q_0=4.0\,\text{GeV}$.  Right: The cross section for $Q_0=20.0\,\text{GeV}$. Note the improvement in the consistency of the three terms, even at $Q=Q_0=4.0\,\text{GeV}$ in the left panels, with respect to the standard approach shown in \fref{tmdg_lines}. As larger $Q_0$ are considered (e.g., with the larger scale $Q_0=20\,\text{GeV}$ above) the three curves begin to converge in the $m \ll \Tsc{q}{} \ll Q_0$ region.
        }
        \label{f.tmdhso_lines}
\end{figure}
\subsection{In a hadron structure oriented approach}
\label{s.hso}

Next, we contrast the conventional approach of the preceding subsection with the HSO steps from Ref.~\cite{Gonzalez-Hernandez:2022ifv} and \srefs{sidis}{largeqtasymptote} of this paper. 

It should be emphasized that the two ``approaches'' being contrasted here refers only to specific phenomenological implementations and not to the basic theoretical setup. The fundamental TMD factorization theorem and the evolution equations are always the standard ones, and they are never modified. What distinguishes the HSO approach to phenomenological implementations from the conventional one is that the former imposes constraints on the input TMD parametrizations that guarantee consistency with collinear factorization in the appropriate limits. To see what this means more clearly, it may be helpful to recall that it is straightforward (though unnecessary) to use the $\bstar$ method to rewrite the HSO expression in \eref{Wtermev0_finalversion} in terms of the $g$-functions defined in \eref{gdef}, but with the explicit HSO parametrizations for $\tilde{f}_{j/p}(\xbj,\T{b}{};\mu_{Q_0},Q_0^2)$ and $\tilde{D}_{h/j}(\zh,\T{b}{};\mu_{Q_0},Q_0^2)$. The final form of the evolved TMD pdfs and ffs are exactly the same. The full set of steps for translating the HSO approach into the conventional one may be found in Sec.~IX of \cite{Gonzalez-Hernandez:2022ifv}.
Cast in this way, the HSO approach is identical to the conventional one except that it imposes additional and important consistency conditions directly on the $g$-functions. 
In the treatment in this paper, this amounts to using \eref{K_param_final}, \eref{candidateqff} and \eref{candidateqpdf} (or, more generally, any other set of parametrizations that arise from the steps in 
Ref.~\cite{Gonzalez-Hernandez:2022ifv}) inside \erefs{evolvedd4}{evolvedd4pdf} instead of the conventionally unconstrained ansatzes like \erefs{gmodel}{gkmodels}. 

We have focused on the kinematics of the $Q \approx Q_0$ region, since the lowest acceptable values of $Q$ are where one typically expects nonperturbative hadron structure effects to be most pronounced, and thus it is where nonperturbative versions of relations like \eref{int_rel_basic} and \eref{cutoff_def} become especially important. 

The steps for calculating the TMD term 
in the HSO approach were reviewed in \srefs{sidis}{models}.
If we specialize to the additive structure in~\sref{pdfsandffs} for the TMD parametrizations, then the HSO approach amounts to simply calculating \eref{Wtermev0_finalversion} with the parametrizations in \eref{candidateqff} and \eref{candidateqpdf}. That is, we use
\begin{align}
\left[ f_{j/p}, D_{h/j} \right] &{}\to
    \int \frac{\diff[2]{\T{b}{}}}{(2 \pi)^2}
    ~ e^{-i\T{q}{}\cdot \T{b}{} }
    ~ \tilde{f}_{j/p}(\xbj,\T{b}{};\mu_{Q_0},\mu_{Q_0}^2) 
    ~ \tilde{D}_{h/j}(\zh,\T{b}{};\mu_{Q_0},\mu_{Q_0}^2)\no&
  \,\times
  \exp\left\{  
        \tilde{K}(\Tsc{b}{};\mu_{Q_0}) \ln \parz{\frac{ Q^2 }{ Q_0^2}}
          +\int_{\mu_{Q_0}}^{\mu_Q}  \frac{ \diff{\mu'} }{ \mu' }
          \biggl[ 2 \gamma(\alpha_s(\mu'); 1) 
                 - \ln\frac{Q^2}{ {\mu'}^2 } \gamma_K(\alpha_s(\mu'))
          \biggr]
  \right\} \, ,
\label{e.replaceWtmd2}
\end{align}
with Eqs.~(\ref{e.F12tmd}). In the replacement, the $\tilde{D}_{h/j}$ and $\tilde{f}_{j/p}$ are now to be understood to be the $\Tsc{b}{}$-space version of
the parametrizations from 
\eref{candidateqff} and \eref{candidateqpdf} substituted into \erefs{evolvedd4}{ev_factor}. Explicit expressions for the input $\Tsc{b}{}$-space TMD functions are listed in \aref{bTspaceTMD}. 
We denote the resulting structure functions by $F^{\text{TMD}}_\text{HSO}$. 
These are the underlined correlation functions from \cite{Gonzalez-Hernandez:2022ifv}\footnote{Actually, these symbols refer to a class of models for the TMD pdfs and ffs since at this stage we still need to specify the exact form of the nonperturbative transverse momentum dependence in $\np{f}{i/p}$ and $\np{D}{h/j}$. We will use the same notation for all calculations that use this general approach.}, or, if we restrict $Q \approx Q_0$ and use the approximation in \eref{Eapprox}, they are just the $\Tsc{b}{}$-space input functions themselves.  
With $\order{\alpha_s}$ perturbative coefficients, their structure is
\begin{align}
\label{e.Ftmdhso}
    F^{\text{TMD}}_{\text{HSO}}\sim{}&\left(|H|^2_j\right)^{\order{\alpha_s}}\,\left[ f_{j/p}, D_{h/j} \right]\,,
\end{align}
where $\left(|H|^2_j\right)^{\order{\alpha_s}}$ is the hard coefficient in \eref{hardpart}, with kinematic factors and sums over flavors. 

For the asymptotic term, we start from $F^{\text{TMD}}_{\text{HSO}}$ and use the $m \ll \Tsc{q}{} \ll Q$ approximation in \eref{add_subtract2} in place of $\left[ f, D \right]$, so that the asymptotic structure functions are
\begin{align}
\label{e.Fasyhso}
    F^{\text{ASY}}_{\text{HSO}}\sim{}&\left(|H|^2_j\right)^{\order{\alpha_s}}\,\left[ f_{j/p}, D_{h/j} \right]_{\text{ASY}}\, .
\end{align}

For calculating the $\order{\alpha_s}$ fixed order structure function in $\Tsc{q}{} \approx Q_0$ collinear factorization (see, for example, Ref.~\cite{Wang:2019bvb}), we use
\begin{align}
\label{e.Ffohso}
{}&\frac{\left(|H|^2_j\right)^{\order{\alpha_s}}}{e^2_j}\,F^{\text{FO}}_{\text{ST}} = F^{\text{FO}}_{\text{ST}} + \order{\alpha_s(\mu_{Q_0})^2} \,,
\end{align}
where $F^{\text{FO}}_{\text{ST}}$ are the $\msbar$ structure functions of \eref{F12fo}. 
Keeping the overall factor 
in $F_\text{ST}^{\text{FO}}$ does not formally change the treatment at the $\order{\alpha_s}$ level, but retaining it improves the agreement with the asymptotic term of \sref{largeqtasymptote} in the $m \ll \Tsc{q}{} \ll Q_0$ limit.

We show numerical examples of $F^{\text{TMD}}_{\text{HSO}}$, $F^{\text{ASY}}_{\text{HSO}}$ and $F^{\text{FO}}$ in \fref{tmdhso_lines}, calculated using the Gaussian models of \eref{npmodels} in \eref{candidateqff} and \eref{candidateqpdf}. The kinematics are the same as in \fref{tmdg_lines}, and the nonperturbative parameters take the values
\begin{equation}
M_{\text{F}}= M_{\text{D}}/z=0.1\,\text{GeV} \, ,
\end{equation}
so that our treatment of the nonperturbative contribution is comparable to the conventional treatment in \fref{tmdg_lines}. Aside from the transition to a tail region, the Gaussian model mimics the power-law behavior of $g$-functions in \eref{gdef} with \eref{gmodel} for the conventional approach. As in \fref{tmdg_lines}, we show the case of a lower input $Q_0 = 4.0$~\text{GeV} in the left panels of \fref{tmdhso_lines}, and a large $Q_0 = 20.0$~\text{GeV} in the right panels. 
The upper two panels show the plots on a logarithmic scale to magnify the improvements at large transverse momentum. To magnify the effect of the improvement on the small transverse momentum region, we have replotted the same graphs on linear vertical axes and over a smaller $\Tsc{q}{}$ range in the lower two panels. The qualitative and quantitative improvements of the HSO over the conventional approach are especially visible on the linear axes.
For these calculations we have used the approximation 
$\overline{Q}_0\to Q_0$ in \erefs{evolvedd4}{evolvedd4pdf}
because this allows us to utilize the analytic expressions for the TMD pdf and ff parametrizations. We confirm in \fref{input_vs_underline}, however, that the effect of the evolution factor is negligible at the input scale. This is by design; the evolution factor is only relevant for evolving to $Q$ well above the input scale. 
\begin{figure}
     \centering
         \includegraphics[width=0.6\textwidth]{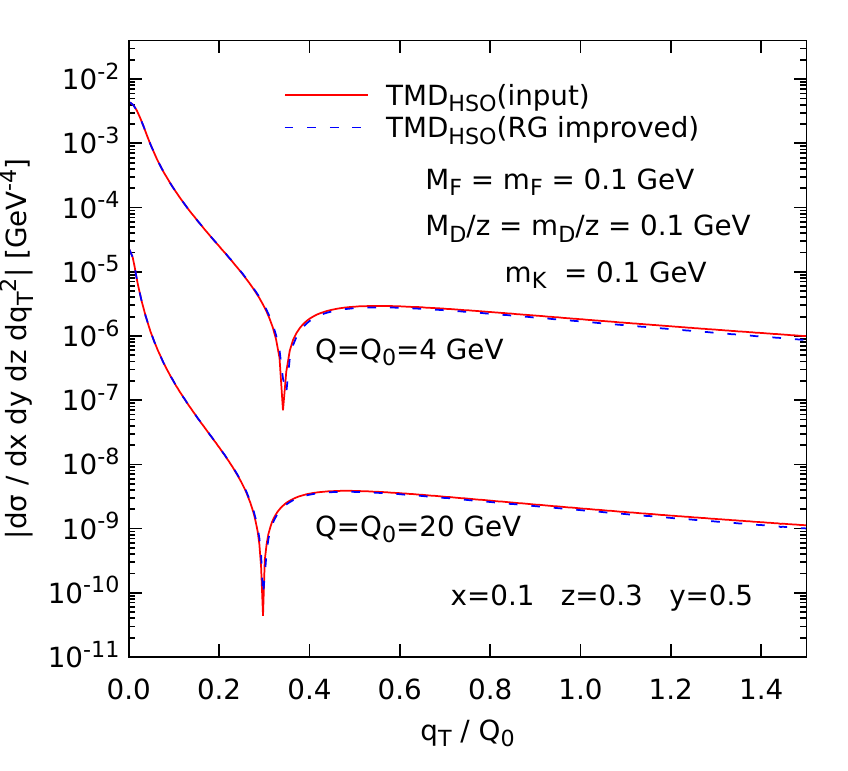}
        \caption{ 
        Comparison of the two versions of the HSO approach discussed in the text, at the two values of the input scale considered, $Q_0=4.0\,\text{GeV}$ and $Q_0=20.0\,\text{GeV}$. The red solid lines show the TMD term calculated directly with the input functions of \eref{candidateqff} and \eref{candidateqpdf}, as it was done in our examples in  \fref{tmdg_lines} and \fref{tmdhso_bands}. The blue dashed lines show the TMD term in the HSO approach but with renormalization group (RG) improvement (the underline version of functions from Ref.~\cite{Gonzalez-Hernandez:2022ifv}) applied at very small $\Tsc{b}{}$, implemented in \erefs{evolvedd4}{ev_factor}, with the transition function $\overline{Q}_0$ of \aref{interp}. In the HSO approach, for $Q\approx Q_0$, these RG improvements affect only the large $\Tsc{q}{}$ region of the  cross section. For our examples in this article, even for $\Tsc{q}{}/Q_0\approx 1.5$, differences are not significant.
        }
        \label{f.input_vs_underline}
\end{figure}

Comparing \fref{tmdhso_lines} with \fref{tmdg_lines} confirms that, in terms of maintaining consistency with the collinear factorization region, there is a very substantial improvement with the HSO approach as compared with the conventional approach. For $Q_0 = 4.0$~GeV, the TMD and asymptotic terms match nearly exactly for all $\Tsc{q}{} \gtrsim 0.5$~GeV up to $Q_0$. There is also a region around $\Tsc{q}{} \approx 0.5$~GeV where all three calculations smoothly overlap. Notice also that the region of overlap becomes better defined when going from the left panel (low input scale) to the right panel (high input scale) of \fref{tmdhso_lines}. 
And, with the larger $Q_0$, the agreement between the TMD and asymptotic terms is nearly exact over the whole visible range of $\Tsc{q}{}$. Thus, the HSO plots exhibit the expected trends when choosing larger or smaller values of $Q_0$. Of course, the calculations with $Q_0$ as large as $20$~GeV are not physically sensible, but they confirm that the two ways of computing the mid-$\Tsc{q}{}$ behavior (with asymptotic and TMD terms) are compatible and consistent in the limit of a fixed $\Tsc{q}{}/Q_0$ ratio and large $Q_0$. 

From \fref{tmdg_bands}, it is clear that in order to correct the large-$\Tsc{q}{}$ behavior of the TMD-term in the conventional methodology to recover the asymptotic term, one would need to make further adjustments to the nonperturbative, non-tail part of the parametrization. But it would have to be done in a way that allows nonperturbative transverse momentum parameter dependence to propagate to unacceptably large $\Tsc{q}{}$. That could be through both explicit nonperturbative parameters like $M_F$ and $M_D$ and through the residual dependence on $\bmax$. In order to reduce the $\bmax$ dependence at large $\Tsc{q}{}$ to acceptable levels while forcing the TMD and asymptotic terms to converge in \fref{tmdg_bands}, one would have to allow dramatic dependence on nonpertubative parameters that affects the behavior at unacceptably large transverse momentum.
To illustrate that the HSO approach addresses this problem, we plot the HSO structure functions, again at the input scale, $Q=Q_0=4.0\,\text{GeV}$, but now with both the Gaussian models of \eref{npmodels} and the spectator diquark models of \eref{spectatorpdf}, and with the same ranges of values of the nonperturbative mass parameters as were used in the connventional treatment. In the HSO approach, there is no $\bstarsc$ or $\bmax$, and the TMD and asymptotic terms converge toward one another automatically. The results are shown in \fref{tmdhso_bands}, with red bands showing the effect of adjusting the nonperturbative mass parameters in the range of \erefs{range1}{range2},
and with the Gaussian model in the left-hand panel and the spectator diquark model in the right-hand panel. In each case, we also display the HSO asymptotic (dashed blue line) and fixed order terms (dot-dashed black line).\footnote{ Since  $F^{\text{ASY}}_{\text{HSO}}$ is calculated with cutoff collinear functions, they also depend on the values of the mass parameters and should in principle be also displayed as bands in \fref{tmdhso_bands}. However, the variations are negligibly small for the ranges of the mass parameters considered here, so for visibility we show only central lines instead.} To see the improvement brought about by the HSO approach, these plots should be compared with the analogous plot in \fref{tmdg_bands} of the conventional treatment. 

The small-$\Tsc{q}{}$ regions in both of the cases shown in \fref{tmdhso_bands} exhibit the behavior of their respective nonpertubative models. As $\Tsc{q}{}$ grows, the red bands around the TMD curves converge around the asymptotic term, until the the TMD and asymptotic curves are indistinguishable, independently of the nonperturbative model or the values used for $M_D$ and $M_F$. This illustrates how the HSO approach enforces a smooth transition to a region that is insensitive to the value of nonperturbative transverse momentum dependence parameters.  Even with the spectator model on the right, where the TMD curves come with visible bands close to the zero node, the curves still match the general shape of the asymptotic term down to $\Tsc{q}{} \approx 0.5\,\text{GeV}$. 
The HSO approach ensures this type of behavior.  

With the Gaussian model in the left-hand panel  \fref{tmdhso_bands}
, the bands show that agreement between the TMD term and the asymptotic term in the region of $\Tsc{q}{} \approx 0.5$ to $\approx 1.0$~GeV requires that the mass parameters be kept rather small. For spectator model, the right-hand plot shows that there is more flexibility to adjust the nonperturbative parameters without spoiling approximate agreement with the asymptotic term at mid $\Tsc{q}{}$. 

In \fref{tmdhso_lines} and \fref{tmdhso_bands}, we also plotted the $\Tsc{q}{} \approx Q$ fixed order curves to show its approximate overlap with the asymptotic and TMD terms in a region of mid $\Tsc{q}{}$. In these calculations, we used $\msbar$ pdfs and ffs. As mentioned in the discussion after \eref{stdef}, it may turn out to be preferable to use the cutoff definitions for the collinear functions to match what is done with the asymptotic term. For the purposes of this paper, however, the difference between the two is small enough to ignore, as can be seen in \fref{Cutoff_vs_MSbar_CollinearPDFsFFs} where we plot the ratios of the collinear pdfs and ffs defined with the cuttoff scheme and the $\msbar$ scheme. For the ranges of $x$ and $z$ that we have consider in this paper, the difference between the schemes is $\lesssim 10\%$, which is comparable to the spread between the asymptotic and fixed order curves in \fref{tmdhso_lines}. It is perhaps interesting that the switch from the $\msbar$ to the cutoff pdfs tends to move the fixed order curve closer to the asymptotic curve. However, we leave the question of whether switching to all cutoff definitions can improve the treatment to future work.

The above style of analysis can be applied directly to the individual TMD correlation functions instead of the full structure functions, and this may be a preferred way to organize the discussion in contexts where understanding the role of hadron structure is the primary goal. In particular, given a nonperturbative treatment of the small $\Tsc{k}{}$ region of a TMD pdf or ff, we may confirm that the TMD function matches its order $\alpha_s^n$ tail at $\Tsc{k}{} \approx Q_0$. An example is shown in \fref{tmdhso_bands_2} for the Gaussian core model. The bands show the effect of varying the mass parameters as in the left panel of \fref{tmdhso_bands}, calculated as in \eref{candidateqff} and \eref{candidateqpdf}. 
The correlation functions are the TMD pdf of up-quarks in a proton (left panel), and the TMD ff of up-quarks into $\pi^+$ (right panel). (These are exactly the functions used in the cross section of \fref{tmdhso_bands}.)  
The dot-dashed lines are the corresponding perturbative calculations in \eref{candidateqff_pert} and \eref{candidateqpdf_pert}. These are the ``aysmptotic terms,'' analogous to the dashed curves in \fref{tmdhso_lines}, but corresponding to the separate TMD correlation functions. The plots show that, regardless of the nonperturbative treatment of small $\Tsc{k}{}$, the TMD correlation functions treated in this way are always consistent with their $\Tsc{k}{} \approx Q_0$ behavior, found in collinear factorization, starting at around $\Tsc{q}{}=1.0\,\text{GeV}$.
The analogous plots for other flavors exhibit similar trends.

\begin{figure}
     \centering
         \includegraphics[width=0.49\textwidth]{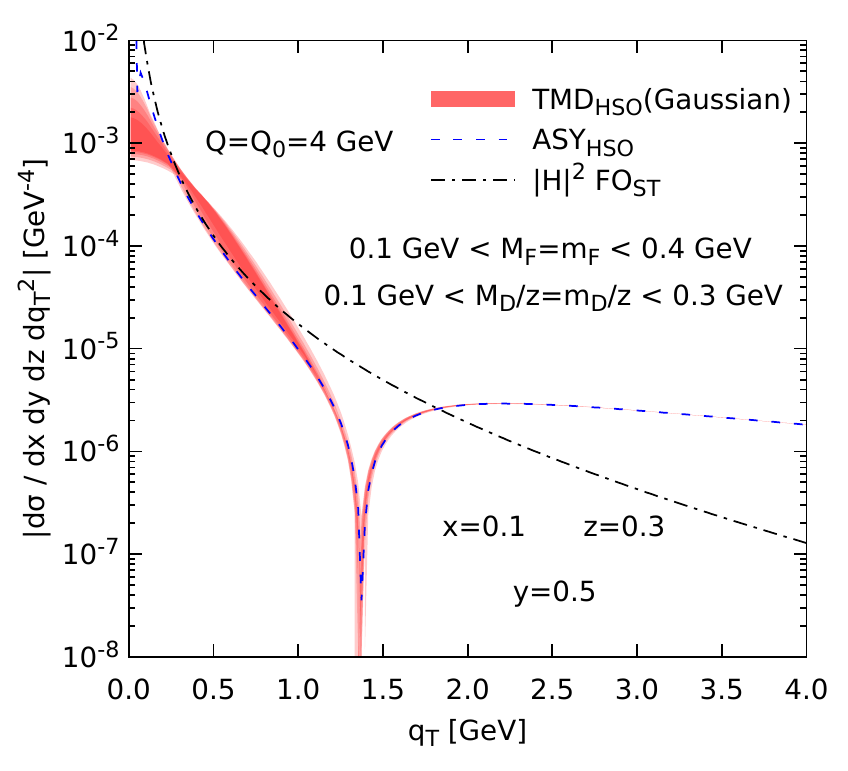}
         \includegraphics[width=0.49\textwidth]{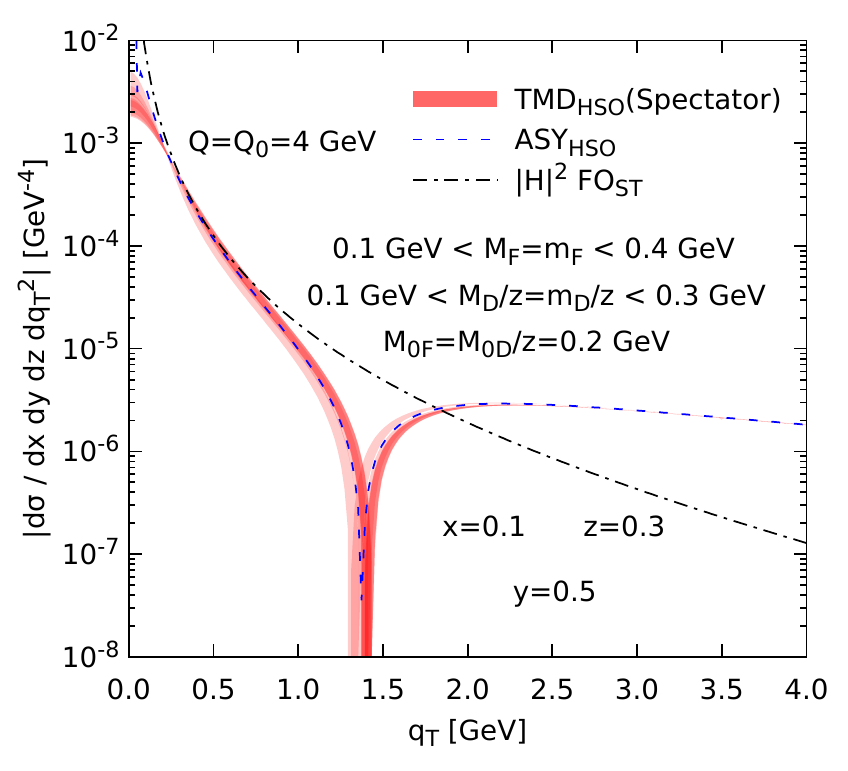}
        \caption{
        Variation of the TMD cross section (absolute value), in the HSO approach, with respect to  small-mass parameters (red bands).  The same kinematics as in the left panel of \fref{tmdg_bands} have been used, so that  $Q=Q_0=4.0\,\text{GeV}$ and no TMD evolution is performed. For visibility, we display only the central lines of the corresponding cross sections with the $F^{\text{ASY}}_{\text{HSO}}$ (dashed blue line) and 
        $F^{\text{FO}}$ as in \eref{Ffohso}
        (dot-dashed black line) approximations, since their variation with the small masses is very mild.
        Left: calculation with the Gaussian ansatz of \eref{npmodels}, obtained by varying the model masses $M_{\text{F}}$ and $M_{\text{D}}$; this is the constrained version of the quadratic models
        for the g-functions of \eref{gmodel}, in the standard approach.  Right:  implementation of the spectator model \erefs{spectatorpdf}{spectatorff} in the HSO approach. In each case, the HSO approach  ensures the consistency of the initial models for TMDs and collinear factorization calculations. Note that our prescription can be readily applied to any other model.
        }
        \label{f.tmdhso_bands}
\end{figure}
\begin{figure}[h!]
\centering
\includegraphics[width=0.7\textwidth]{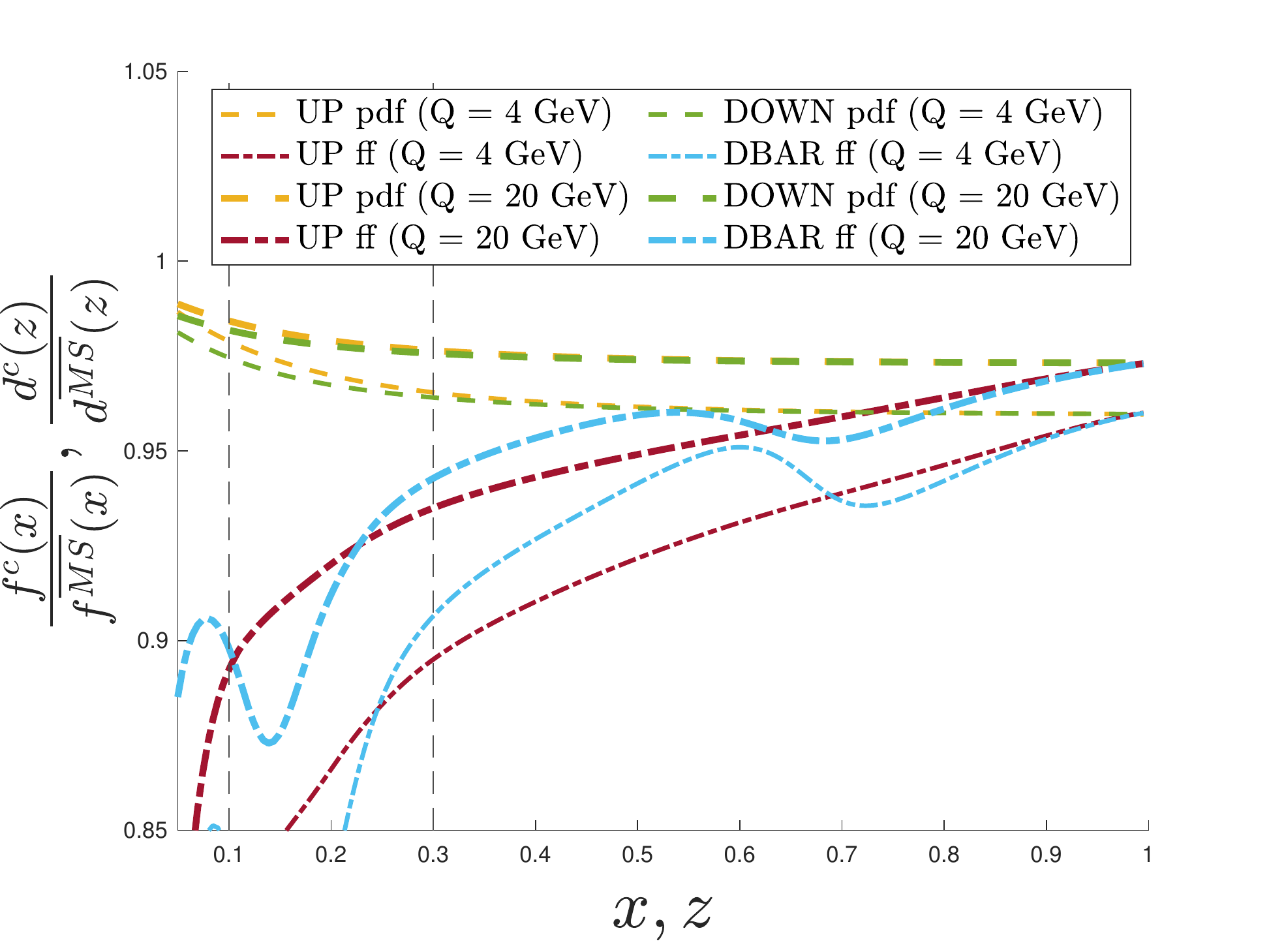}
\caption{Ratio of the cutoff definitions of both collinear pdfs (dashed) and ffs (dot-dashed) and their $\msbar$ counterparts. The results are shown for two values of the hard scale $Q \in \lbrace4, 20\rbrace$ GeV. Only the flavors that contribute most to the cross section under consideration have been shown to facilitate the readability of the plot. The cutoff has been chosen to match the hard scale, i.e. $k_c = Q$ with the choice of the Gaussian model for TMDs  for the ``core" parametrization of both pdfs and ffs. Notice that any other choices for the ``core" model or the nonperturbative mass parameters would only affect the result with power suppressed contributions which, if neglected, make the difference between the two schemes perturbatively calculable as in the last term in \eref{C_def} (ff) and \eref{C_def_pdf} (pdf). The dashed black lines correspond to the choices of $\xbj = 0.1$ and $z = 0.3$ made in our computations throughout the paper. Replacing the two definitions thus accounts for a difference of order $\sim 3\%$ ($Q = 4$~GeV) to $\sim 2\%$ ($Q = 20$~GeV) for the pdf case and $\sim 10\%$ ($Q = 4$~GeV) to $\sim 5\%$ ($Q = 20$~GeV) for the ff case.}
\label{f.Cutoff_vs_MSbar_CollinearPDFsFFs}
\end{figure}
\begin{figure}
     \centering
     \includegraphics[width=0.49\textwidth]{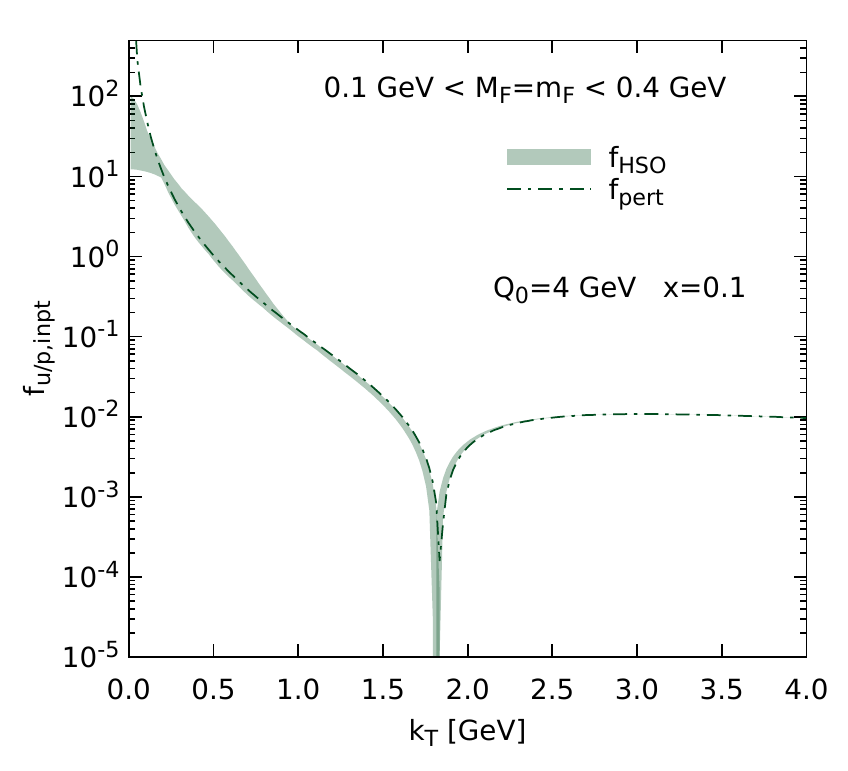}
     \includegraphics[width=0.49\textwidth]{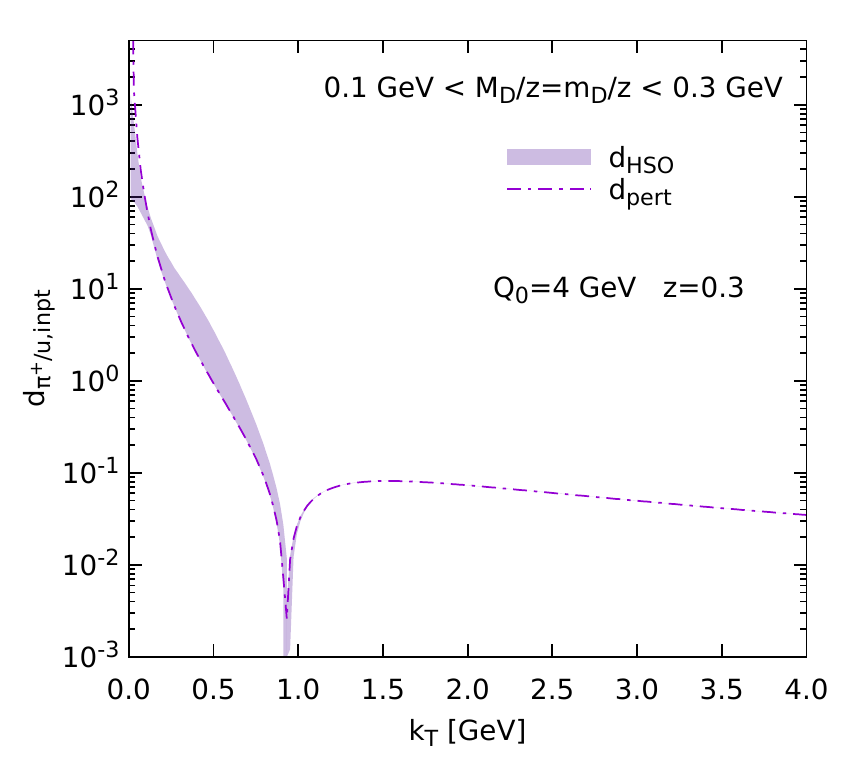}
        \caption{
        Comparison of the TMD functions in the HSO approach and their large-$\Tsc{k}{}$ behaviour predicted by pQCD. The bands show the variation of the TMD pdf of \eref{candidateqpdf} (left) and the TMD ff \eref{candidateqff} (right), with respect to mass parameters, using the Gaussian ansatzes of \eref{npmodels}. The range of masses indicated in the labels are the same as those used to obtained the red band in the left panel of \fref{tmdhso_bands}. The dot-dashed lines show the pQCD  calculation of \eref{candidateqpdf_pert} for the TMD pdf (left), and that of \eref{candidateqff_pert}   for the TMD ff (right). The correct behaviour for the models, has been imposed from the onset in \eref{candidateqff_pert} \eref{candidateqpdf_pert}, through the $A$ and $B$ coefficients. This is indeed a necessary condition for the agreement of the TMD cross section and the asymptotic term in the left panel of \fref{tmdhso_bands}.
        }
        \label{f.tmdhso_bands_2}
\end{figure}

\section{Conclusion}
\label{s.conclusion}

Let us conclude by summarizing the primary results of the last section: 
We have shown how to implement TMD factorization to calculate unpolarized SIDIS cross sections at an input scale $Q_0$ in a way that centers the role of nonperturbative calculations of hadron structure, and we have shown how this leads to a dramatic improvement in the consistency between TMD and collinear factorization, particularly near the input scale $Q_0$. 
Our approach, which we have called a ``hadron structure oriented'' approach in this paper, and which is based upon the setup in~\cite{Gonzalez-Hernandez:2022ifv}, imposes additional constraints beyond what is standard in the more conventional style of implementing TMD factorization, reviewed above in \sref{conventional}. These extra constraints are designed especially to preserve a TMD parton model interpretation (in the sense of preserving \eref{int_rel_basic}) for small transverse momentum behavior while ensuring a consistent transition to collinear factorization at $\Tsc{q}{} \approx Q_0$ and $Q_0 \gg \Lambda_\text{QCD}$. 
We have emphasized throughout that it is straightforward to  
swap the parametrization of the nonperturbative core of a TMD pdf or ff in the HSO approach, so that any preferred model or nonperturbative technique for describing the small transverse momentum region may easily be incorporated into future implementations. We highlighted this modular feature of the HSO approach by exchanging a Gaussian model for a spectator diquark model in \fref{tmdhso_bands}; replacing one description of the nonperturbative core by another leaves the $\Tsc{q}{} \approx Q_0$ region of the TMD term unaffected and consistent with large-$\Tsc{q}{}$ collinear factorization. 

Of course, there are still other open questions with regard to the domain of applicability of TMD factorization to processes like SIDIS. For example, a definitive lowest value for $Q$ (for each $\xbj$ and $z$) in SIDIS below which TMD factorization techniques absolutely cease to be useful remains to be determined. It is likely that a sharp transition does not exist. A related question is that of how high $\Tsc{q}{}$ may become before the TMD term alone is no longer sufficient, and the description must transitions into a $\Tsc{q}{} \approx Q$ region where TMD factorization fails and one must rely entirely on fixed order collinear factorization. (This is the issue of the ``$Y$-term'' alluded to in the introduction.) Below some numerical value of $Q$, it is no longer meaningful to separate a cross section into distinct large ($\Tsc{q}{} \approx Q$) and small  ($\Tsc{q}{} \approx \Lambda_\text{QCD}$) transverse momentum regions. These should probably be viewed as open empirical questions, to be confronted by future experimental tests. But posing them in a clear way requires unambiguous and internally consistent steps like those we have described here and in \cite{Gonzalez-Hernandez:2022ifv}  with the HSO approach. 

A separate phenomenological issue is that one generally finds tension between data for large transverse momentum in processes like SIDIS and Drell-Yan scattering and calculations performed with existing collinear pdf and ff fits~\cite{Daleo:2004pn,Gonzalez-Hernandez:2018ipj,Bacchetta:2019tcu,Wang:2019bvb}.
This suggests that it will be important for future phenomenological  efforts to fit TMD and collinear functions simultaneously in a full TMD factorization context. Of course, for this to be meaningful the nonperturbative parts need to be combined with collinear factorization in a consistent procedure, and this is what the HSO approach is meant to provide. 

Extending the treatment in this paper of SIDIS to other processes like Drell-Yan scattering is straightforward. Moreover, order $\alpha_s^2$ and even $\alpha_s^3$ versions of the parametrizations are obtainable from straightforward, albeit somewhat cumbersome, translations of existing results. It will ultimately be necessary as well to formulate the spin and azimuthal dependent observables in TMD factorization in a manner analogous to what we have done here for the unpolarized case. There, interesting subtleties arise from matches and mismatches between small and large transverse regions of the TMD pdfs and ffs~\cite{Bacchetta:2008xw,Qiu:2020oqr,Rogers:2020tfs}. 
In addition, there exist other QCD formalisms that invoke the notion of a TMD or unintegrated parton density and find complications with preserving relationships like \eref{int_rel_basic}, see for example Refs.~\cite{Kimber:2001sc,Watt:2003mx} and the discussion in Refs.~\cite{Guiot:2022psv,Golec-Biernat:2018hqo,Guiot:2019vsm}. We hope that our work might provide some input in resolving these problems. 
Finally, it bears mentioning that the HSO approach that we advocate here is entirely compatible with other frameworks for setting up TMD factorization and/or transverse momentum resummation methods, including soft-collinear effective theory based approaches \cite{Catani:2000vq,Catani:2012qa,Camarda:2019zyx,Becher:2006mr,Becher:2010tm,GarciaEchevarria:2011rb}.  

For our next steps, we plan to perform explicit phenomenological extractions within the HSO approach discussed here. It has the advantage of placing us in a position to systematically analyze the contributions from any nonperturbative models (e.g., the spectator model) for the small transverse momentum region separately from the large transverse momentum perturbative tails. Such analyses can then be related directly to specific regions of observable transverse momentum in experimental data, in the spirit of, for example, the discussion of Fig.~17 in \cite{COMPASS:2017mvk}. Ultimately, one hopes to infer, from the extracted correlation functions, information about the underlying nonperturbative physics. To see an example of where this will be useful, consider Ref.~\cite{Schweitzer:2012hh},  which describes a treatment of intrinsic transverse momentum in a field theoretic chiral constituent quark model where the chiral symmetry breaking scale is large relative to the constituent quark mass. The HSO approach discussed in this paper is ideally suited for connecting this and similar descriptions to SIDIS data in the context of a complete TMD factorization treatment. Notice in particular that the additive model we constructed in \srefs{pdfsandffs}{models} aligns naturally with the Gaussian-plus-tail type of description in Ref.~\cite{Schweitzer:2012hh}. 
More generally, adopting an HSO approach enables us to begin to ask more specific and detailed phenomenological questions about the adequacy of specific theories of nonperturbative small transverse momentum behavior.  

The elements necessary for these and other studies designed to identify separate perturbative and nonperturbative structures are in place now, and extensions to higher orders in $\alpha_s$ are straightforward, given existing results in the literature. 

\vskip 0.3in
\acknowledgments
We thank Fatma~Aslan, Mariaelena~Boglione, Nobuo~Sato, and Andrea~Simonelli for useful conversations. J.O.~Gonzalez-Hernandez acknowledges funding from the European
Union’s Horizon 2020 research and innovation programme under grant agreement No 824093.
T.~Rogers and T.~Rainaldi were supported by the U.S. Department of Energy, Office of Science, Office of Nuclear Physics, under Award Number DE-SC0018106. 
This work 
was also supported by the DOE Contract No. DE- AC05-06OR23177, under which 
Jefferson Science Associates, LLC operates Jefferson Lab. 


\begin{appendix}

\section{Scale transformation function}
\label{a.interp}

The scale transition function in \eref{ev_factor} is in principle entirely arbitrary, see the discussion in Sec.~V of \cite{Gonzalez-Hernandez:2022ifv}, provided it has the general feature that it transitions from $\sim 1/\Tsc{b}{}$  behavior to $Q_0$ at a $\Tsc{b}{}$ slightly below $1/Q_0$. This ensures, by construction, that we avoid modifying the input scale treatment of \eref{Wtermev0_finalversion} in the $Q \approx Q_0$ region. In this paper, namely in \fref{input_vs_underline}, we have adopted the same choice as in Appendix C of~\cite{Gonzalez-Hernandez:2022ifv},  
\begin{align}
\label{e.qbar_param_a}
&\overline{Q}_0(\Tsc{b}{},a) = Q_0 \left[ 1 - \parz{1 - \frac{C_1}{Q_0 \Tsc{b}{}}}
e^{-\Tscsq{b}{} a^2} \right] \, .
\end{align}
The constant $C_1$ has the usual numerical value of $C_1 = 2 e^{-\gamma_E} \approx 1.123$. The specific value of $a$ used in \fref{input_vs_underline} is $a=Q_0$. 
\section{TMD parametrization in $\Tsc{b}{}$ space at the input scale}
\label{a.bTspaceTMD}

Here we list the $\Tsc{b}{}$-space versions of \eref{candidateqff} and \eref{candidateqpdf} 
\begin{align}
z^2\inptp{\tilde{D}}{h/j}(\zh, \T{b}{};\mu_{Q_0},Q_0^2) &{}=   K_0\left(\Tsc{b}{}m_{D_{h,j}}\right) \left[A^D_{h/j}(\zh;\mu_{Q_0}) + B^D_{h/j}(\zh;\mu_{Q_0}) \ln\left( \frac{\Tsc{b}{}Q_0^2 e^{\gamma_E}}{2 m_{D_{h,j}}} \right) \right] \no
&{}+K_0\left(\Tsc{b}{}m_{D_{h,j}}\right) A^{D,g}_{h/j}(\zh;\mu_{Q_0})   \no
&{} 
+
C^D_{h/j} \,z^2\np{\tilde{D}}{h/j}(\zh, \T{b}{};Q_0^2) \, , 
\label{e.candidateqffbTspace}
\end{align}
\begin{align}
\inptp{\tilde{f}}{i/p}(\xbj,\T{b}{};\mu_{Q_0},Q_0^2) &{}= 
 K_0\left(\Tsc{b}{}m_{f_{i,p}}\right)\left[A^f_{i/p}(\xbj;\mu_{Q_0}) + B^f_{i/p}(\xbj;\mu_{Q_0}) \ln\left( \frac{\Tsc{b}{}Q_0^2 e^{\gamma_E}}{2 m_{f_{i,p}}} \right)\right] \no
&{}+   K_0\left(\Tsc{b}{}m_{f_{g,p}}\right)A^f_{g/p}(\xbj;\mu_{Q_0}) \no
&{} 
+
C^f_{i/p} \,\np{\tilde{f}}{i/p}(\xbj,\T{b}{};Q_0^2) \, ,
\label{e.candidateqpdfbTspace}
\end{align}
where
\begin{align}
    \np{\tilde{f}}{i/p}(\xbj,\T{b}{};Q_0^2) &= \int\diff{^2}{\T{k}{}}e^{-i\T{k}{}\cdot \T{b}{}}\np{f}{i/p}(\xbj,\T{k}{};Q_0^2) \, , \\
     \np{\tilde{D}}{h/j}(\xbj,\T{b}{};Q_0^2) &= \int\diff{^2}{\T{k}{}}e^{i\T{k}{}\cdot \T{b}{}}\np{D}{h/j}(\xbj,\zh\T{k}{};Q_0^2) \, ,
\end{align}
which for the Gaussian and spectator model that we use read
\begin{align}
\label{e.npmodelsbT}
\np{\tilde{f}}{i/p}^{\text{Gauss}}(\xbj,\T{b}{};Q_0^2)={}&
e^{-\frac{\Tscsq{b}{}M_\text{F}^2}{4}}\,,
\\
z^2\np{\tilde{D}}{h/j}^{\text{Gauss}}(\zh, \T{b}{};Q_0^2)={}&
e^{-\frac{\Tscsq{b}{}M_\text{D}^2}{4 z^2}}\, ,\\
\np{\tilde{f}}{i/p}^{\text{Spect}}(\xbj,\T{b}{};Q_0^2)={}&
\frac{\mospec^2 \left(\Tsc{b}{} \mospec\right){}^2}{4 \left(2 \mspec^2+\mospec^2\right)} \left(6 K_2\left(\Tsc{b}{} \mospec\right)+\frac{\left(\mspec^2-\mospec^2\right) \left(\Tsc{b}{} \mospec\right)}{\mospec^2} K_3\left(\Tsc{b}{} \mospec\right)\right)
\,,
\\
 z^2\np{\tilde{D}}{h/j}^{\text{Spect}}(\zh, \T{b}{};Q_0^2)={}&
\frac{\mospecff^2 \left(\Tsc{b}{} \mospecff\right)}{2 z^2 \left(\mspecff^2+\mospecff^2\right)} \left(4 z K_1\left(\frac{\Tsc{b}{} \mospecff}{z}\right)+\frac{\left(\mspecff^2-\mospecff^2\right) \left(\Tsc{b}{} \mospecff\right)}{\mospecff^2} K_2\left(\frac{\Tsc{b}{} \mospecff}{z}\right)\right).
\end{align}

\end{appendix}
\bibliography{bibliography}

\end{document}